\documentclass[twocolumn,epjc3]{svjour3}

\usepackage{amsmath,amssymb}
\usepackage{dsfont}
\usepackage{graphicx} 
\usepackage{epstopdf} 
\usepackage{slashed}
\usepackage{enumerate}

\usepackage{subfigure}
\usepackage{color}
\usepackage{multirow}
\usepackage[scientific-notation=true]{siunitx}
\usepackage{hyperref}
\usepackage{caption,cite}

\usepackage{footmisc}

\usepackage{array}
\newcolumntype{L}[1]{>{\raggedright\let\newline\\\arraybackslash\hspace{0pt}}m{#1}}
\newcolumntype{C}[1]{>{\centering\let\newline\\\arraybackslash\hspace{0pt}}m{#1}}
\newcolumntype{R}[1]{>{\raggedleft\let\newline\\\arraybackslash\hspace{0pt}}m{#1}}

\numberwithin{equation}{section}

\newcommand{\be}{\begin{eqnarray*}}
\newcommand{\ee}{\end{eqnarray*}}

\newcommand{\bee}{\begin{eqnarray}}
\newcommand{\eee}{\end{eqnarray}}
\newcommand{\beeq}{\begin{equation}}
\newcommand{\eeeq}{\end{equation}}

\begin{document}

\journalname{}

\title{Constraining new resonant physics with top spin polarisation information}
\author{Christoph Englert\thanksref{e1,addr1}
	\and James Ferrando\thanksref{e2,addr2}
	\and Karl Nordstr\"om\thanksref{e4,addr1}}
\thankstext{e1}{E-mail: christoph.englert@glasgow.ac.uk}
\thankstext{e2}{E-mail: james.ferrando@desy.de}
\thankstext{e4}{E-mail: k.nordstrom.1@research.gla.ac.uk}
\institute{SUPA, School of Physics and Astronomy, University of Glasgow, Glasgow, G12 8QQ, United Kingdom \label{addr1}
 \and
DESY Hamburg, Notkestrasse 85, D-22607 Hamburg, Germany \label{addr2}
}

\date{\today}

\maketitle

\begin{abstract}
We provide a comprehensive analysis of the power of including top quark-polarisation information to kinematically challenging $t\bar t$ resonance searches, for which ATLAS and CMS start losing sensitivity. Following the general modeling and analysis strategies pursued by the experiments, we analyse the semi-leptonic and the di-lepton $t\bar t$ channels and show that including polarisation information can lead to large improvements in the limit setting procedures with large data sets. This will allow us to set limits for parameter choices where sensitivity from $m(t\bar t)$ is not sufficient. This highlights the importance of spin observables as part of a more comprehensive set of observables to gain sensitivity to BSM resonance searches.
\end{abstract}

\section{Introduction}
Given the lack of any conclusive hint for new physics beyond the Standard Model (BSM), it is important to enhance the sensitivity of collider searches that target new states and interactions that are kinematically accessible at the Large Hadron Collider~(LHC) after the first runs.

Observables which directly reflect the final state momentum transfer, such as invariant mass or transverse momentum distributions are obvious choices for searches for new resonant states. However, if the new physics production cross section is small, these observables might not have enough discriminating power to isolate the signal from the competing backgrounds satisfactorily. In these circumstances, the LHC experiments typically favor multivariate techniques over rectangular cut flows. While this approach can increase the sensitivity dramatically, care needs to be taken during the training stage of the analysis. In particular,  experimental constraints (such as the detector's granularity, response effects etc.) need to be included and understood precisely in order to formulate a realistic sensitivity estimate. Therefore, the reliability of these methods entirely lies within the remit of the expertise of the experimental community.

From a theoretical perspective, in case of low expected BSM cross section, there is still motivation to ask whether observables which are complementary to invariant mass distributions provide sensitivity improvements.

For instance, constraints on the production cross section of new resonant states derived from mass resonance searches are strongly dependent on the assumed width of the new state. As the width gets larger, e.g. in strongly-coupled scenarios, the signal gets increasingly washed out and it becomes more difficult to separate its shape from the smoothly falling background even though the cross section might still be sizable. We will show that spin polarisation observables are precisely observables which can improve the limit setting in such a case.

Assuming large statistics, multi-dimensional analyses in more than one observable become possible. This opens up the opportunity to study a variety of distributions and their correlations. In particular a spin-assisted $t\bar t$ invariant mass search, which is the focus of this work, becomes possible.

Models, which are typically employed by the ATLAS and CMS collaborations to look for and constrain the presence of new resonances are extra dimension scenarios, see~e.g.~\cite{Aad:2015fna,Khachatryan:2015sma}. In particular, the compactified Randall-Sundrum (RS) model of Ref.~\cite{Randall:1999ee} introduces a series of isolated graviton resonances into the 4D effective theory. If SM fields propagate in the entire five-dimensional Anti-de Sitter (AdS) background geometry, the 4D theory will also contain Kaluza-Klein copies of the low energy states that are identified with the SM.

The recent experimental study in~\cite{Aad:2015fna} demonstrated that the constraint on the production cross section of e.g. a 3 TeV gluon $g_{KK}$ decaying to $t \bar{t}$ weakens by almost an order of magnitude when going from $\Gamma/m$ = 10\% to $\Gamma/m$ = 40\%. Such large widths can be problematic from a modeling perspective but are not unexpected in strongly-coupled theories inherent to the dual formulation of RS-type theories.
From this AdS/CFT~\cite{Maldacena:1997re,Witten:1998qj,ArkaniHamed:2000ds,Rattazzi:2000hs} perspective, the top quark being the heaviest particle discovered so far plays a special role as its mass could be direct evidence of (at least partial) compositeness. A potential composite structure of extra resonances could therefore be reflected in the analysis of the associated top quark spin observables, while a $t\bar t$ bump search alone does not access this level of detail.

These BSM-induced effects can be contrasted with the fact that $t \bar{t}$ production in the SM at the LHC is dominated by parity-invariant QCD processes. We therefore can expect to produce a roughly even number of left and right-handed tops in the absence of any BSM physics. At the high invariant masses we consider there is a sizeable contribution from weak processes which makes the SM expectation slightly left-handed. This fact has inspired many studies of top polarisation as a probe into BSM physics, both in pair~\cite{Hikasa:1999wy,Li:2006he,Godbole:2006eb,Godbole:2006tq,Krohn:2011tw,Cerrito:2016qig} and single~\cite{Gajdosik:2004ed,Perelstein:2008zt,Godbole:2011vw,Belanger:2012tm,Aguilar-Saavedra:2017nik} production. As the decays of Kaluza-Klein gluons $g_{KK}$ and gravitons $G_{KK}$ are dominated by right-handed tops these distributions are modified as pointed out in for example~\cite{Agashe:2006hk,Lillie:2007yh}.

The crucial point for including spin information to the limit setting is that increasing the width of a parent particle only has a modest effect on spin observables of its decay products. Therefore, they offer a great opportunity to not only give us more information generically, but also reduce the impact of considering wider signal models. We will show that this allows to enhance the sensitivity of analyses like~\cite{Aad:2015fna}.

Therefore, we consider $pp \to g_{KK}/G_{KK}\to t_R \bar{t}_R$ production in this paper and study both the semi-lep\-tonic and di-leptonic final states of the top decays in the region where the reported sensitivity is low. Our goal is to determine to what extent top polarisation and spin correlation measurements allow us to make stronger empirical statements for the models studied in e.g.~\cite{Aad:2015fna}.\footnote{While our search focuses specifically on the Randall-Sundrum model as it allows us to investigate the interplay of spin observables and cross sections in a theoretically meaningful way they directly generalise to a $Z'$ case with chiral couplings to 3rd generation fermions.} Our results can be considered as a litmus test that motivates the consideration of such observables to the aforementioned multivariate techniques pursued by the experiments.

The paper is organised as follows: in Sec.~\ref{sec:model} we quickly introduce the model and discuss relevant parameter for our analysis to make this paper self-consistent. In Sec.~\ref{sec:semilep} we discuss the semi-leptonic final state, while Sec.~\ref{sec:dilep} focuses on the di-leptonic final state. In Sec.~\ref{sec:results} we summarise our results and present our conclusions in Sec.~\ref{sec:conclusions}.

\section{The Model}
\label{sec:model}
In RS1 models~\cite{Randall:1999ee} the hierarchy problem is solved by introducing an extra compactified dimension $r_{\textrm{UV}} < z < r_{\textrm{TeV}}$ with a warped anti-de Sitter geometry AdS$_5$. This explains fine-tuning in $M_\textrm{Planck}/M_\textrm{Weak}$ in terms of the localisation of the 4D graviton near the "Planck" brane, $z=r_{\textrm{UV}}$ with a fundamental scale of $M_\textrm{Planck}$ and the Higgs sector near the "TeV" brane, $z=r_{\textrm{TeV}}$, with a fundamental scale of $M_\textrm{Weak}$. Thanks to the warped geometry we then expect $M_\textrm{Planck}/M_\textrm{Weak} \sim \exp\{\pi k (r_{\textrm{TeV}}-r_{\textrm{UV}})\}$, where $k$ is the AdS curvature scale and $r_C=r_{\textrm{TeV}}-r_{\textrm{UV}}$ is the size of the extra dimension. This is solved by $k r_C \sim 11$ for the observed values of the Planck and weak scales, and hence massively reduces the required fine-tuning. Methods to stabilise the geometry are known~\cite{Goldberger:1999uk}.

\begin{figure*}[t!]
\centering
\subfigure[Di-leptonic $t\bar t$ channel.\label{fig:ttbarmassdil}]{\includegraphics[width=0.45\textwidth]{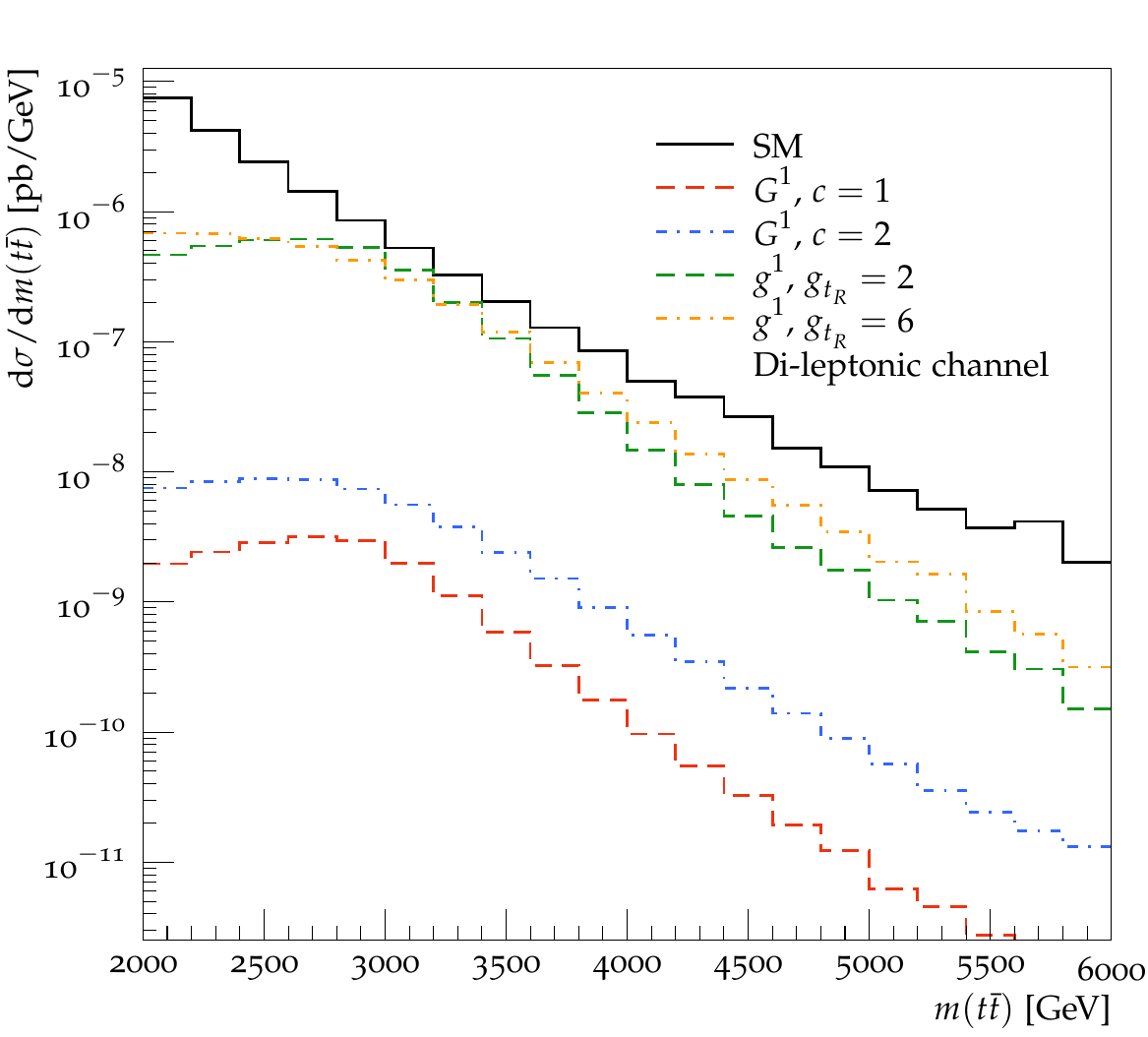}}
\hfill
\subfigure[Semi-leptonic $t\bar t$ channel.\label{fig:ttbarmasssl}]{\includegraphics[width=0.45\textwidth]{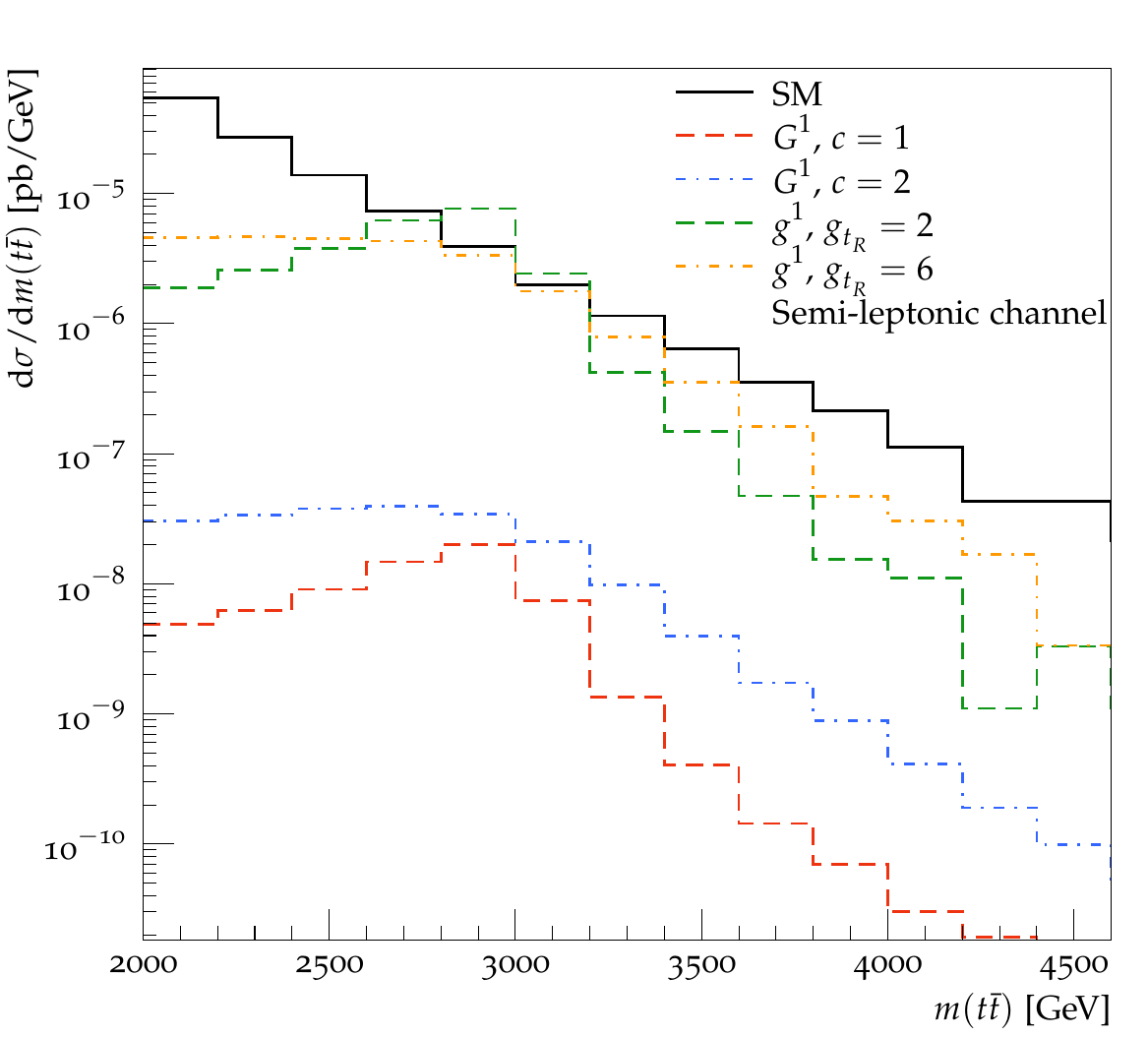}}
\caption{Distributions of $m(t \bar{t})$ for the di-leptonic (a) and semi-leptonic (b) analyses for the background SM $t\bar{t}$ and signal samples.}
\label{fig:ttbarmass}
\end{figure*}

If the SM fermions propagate in all five dimensions, we can additionally explain the structure of the Yukawa sector through localisation~\cite{Gherghetta:2000qt}. The profile of the fermions' wave function is determined by a localisation factor $\nu$ (see~\cite{Lillie:2007yh} for details) which exponentially peaks towards the Planck brane for $\nu < -1/2$ and towards the TeV brane for $\nu > -1/2$ (this can be understood as mixing with CFT bound state in the dual picture, see~\cite{ArkaniHamed:2000ds,Contino:2004vy} for details). To avoid constraints from $Z \to b_L \bar b_L$ while reproducing the correct Yukawa structure we will gauge right handed isospin and set $\nu_{t_R} > \nu_{Q3_L} > \nu_{\textrm{other}}$ following~\cite{Agashe:2003zs}. In general we will keep $\nu_{\textrm{other}} < -1/2$.

Setups with the right-handed top quark localised close to the TeV brane, a flat third generation left-handed quark doublet profile, and other the fermions localised close to the Planck brane are phenomenologically viable~\cite{Agashe:2003zs}. Thanks to $t_R$ living on the TeV brane and $(t,b)_L$ being almost flat, the dominant decay mode of $g_{KK}$ and $G_{KK}$ is to $t_R \bar{t}_R$.

These are typical parameter choices that underpin the experimental analyses. For the graviton, branching fractions to $hh$ and $V_L V_L^\dagger$ are also sizeable as the Higgs and therefore also the longitudinal modes of the weak bosons are located on the TeV brane, but strong constraints on the masses of both particles $m(g_{KK})$ and $m(G_{KK})$ are typically derived from top resonance sear\-ches \cite{Aad:2015fna,Khachatryan:2015sma}.

Our model setup follows these strategies of ATLAS and CMS~\cite{Aad:2015fna,Khachatryan:2015sma} but varies slightly between the gluon and graviton signals. In general the gluon will always be easier to detect due to much larger cross sections as it can be produced efficiently through $u \bar u$ and $d \bar d$ annihilation, whereas graviton production is dominated by gluon fusion. As such it does not make sense to compare identical parameter points and we focus on choices which give a (relatively) narrow and a wide resonance for each signal model.

For our graviton samples we consider the above extreme case where $t_R$ is localised on the TeV brane (i.e. being fully composite), $Q3_L$ is very close to flat, and the decay widths of the lightest KK graviton resonance therefore are:
\begin{align}
  &\Gamma(G^1 \to t_R \bar t_R) = 9 \frac{(3.83 c)^2 m_{G^1}}{960 \pi}\,, \\
  &\Gamma(G^1 \to \phi \phi) = 4 \frac{(3.83 c)^2 m_{G^1}}{960 \pi}\,,
\end{align}
with $c = k/M_\textrm{Planck}$. The factor of $3.83$ is the first root of the Bessel function $J_1$ which is encountered in RS models for the wave function along the compactified direction, and which stems from the boundary condition for gravitons. $\phi$ sums over $Z_L$, $W_L$, and $h$. Decays to right-handed tops are therefore dominant at $\sim 70\%$ and offer good prospects for detection, however, both $ZZ$~\cite{Agashe:2007zd} and $WW$ searches offer additional information (see \cite{Aaboud:2016okv,Sirunyan:2016cao}). We consider two values of $c = \{1, 2\}$ which correspond to widths of $\Gamma_{G^1}/m_{G^1} = \{6.2\%,25\%\}$.

For our gluon sample we soften the localisation requirement and set $\nu_{Q3_L} \sim -0.4$ and vary $\nu_{t_R} \sim \{-0.3, 0\}$ which corresponds to effective couplings of $g_{g^1 b_L \bar b_L} = g_{g^1 t_L \bar t_L} = g_S$, and $g_{g^1 t_R \bar t_R} = \{2, 6\}g_S$. These give widths of $\Gamma_{g^1}/m_{g^1} = \{6.2\%,37.5\%\}$ and branching ratios to $t \bar t = \{78.5\%, 96.5\%\}$. While always dominated by right-handed tops, the fraction of right-handed to left-handed tops also changes which should be reflected in the polarisation observables.

\subsection{Event Generation and Analysis}
Our background is leading order semi- and di-leptonic $t \bar{t}$ samples generated using \textsc{MadGraph} 5~\cite{Alwall:2011uj,Alwall:2014hca} and reweighted to the NNLO cross section given in~\cite{Czakon:2012zr,Czakon:2012pz,Czakon:2013goa}. We focus on $\sqrt{s}=14~\text{TeV}$ collisions. Our signal samples are also generated with \textsc{MadGraph} using the UFO model format~\cite{Degrande:2011ua} to import models implemented in the \textsc{FeynRules}~\cite{Alloul:2013bka} language.
These parton level samples are then showered in \textsc{Herwig 7.0.3}~\cite{Bahr:2008pv,Bellm:2015jjp} and analysed using the Rivet framework~\cite{Buckley:2010ar} which we also use for applying smearing and efficiencies to the physics objects according to typical ATLAS Run 2 resolutions (where available, with Run 1 resolutions used otherwise)~\cite{Aad:2012re,ATL-PHYS-PUB-2015-041,Aad:2016jkr} at the beginning of the analysis routine.

\section{Analyses}
\subsection{Semi-leptonic study}
\label{sec:semilep}
\subsubsection{Analysis Selections and Reconstruction}
\label{sec:semilep-analysis}

The analysis of the semi-leptonic samples focuses on reducing non-$t \bar{t}$ backgrounds and reconstructing the individual tops, largely following the boosted approach detailed in~\cite{Aad:2015fna}. We start by finding electrons with $p_T > 25$ GeV for $|\eta| < 2.47$ and muons with $p_T > 25$ GeV with $|\eta| < 2.7$. We then cluster narrow anti-$k_T$~\cite{Cacciari:2008gp} $R=0.4$ jets with $p_T > 25$ GeV inside $|\eta| < 2.8$ and fat Cambridge-Aachen~\cite{Dokshitzer:1997in,Wobisch:1998wt} $R=1.2$ jets with $p_T > 250$ GeV inside $|\eta| < 2$, and require at least one of each after removing narrow jets which overlap with the leading fat jet.

Since we are interested in highly boosted tops, we have to accept some overlap between the lepton and $b$-jet on the leptonic side so we do not require these to be isolated and assume we can veto events with hard leptons from heavy flavour decays inside QCD-produced jets.\footnote{See~\cite{Rehermann:2010vq} for a proof-of-principle investigation using the muon final state.} Following~\cite{Plehn:2011tf}, we top-tag the leading fat jet with \textsc{HEPTopTagger}~\cite{Plehn:2009rk,Kasieczka:2015jma} with a mass drop threshold of 0.8, max subjet mass of 30 GeV, $R_\textrm{filt} = 0.3$, $n_\textrm{filt} = 5$, and $f_W = 0.15$. We require the candidate to have a mass between 140 and 210 GeV and a $p_T > 250$ GeV to be consistent with a boosted top quark. This provides our hadronic top candidate and we require at least one of the narrow jets to be $b$-tagged with an efficiency of 70\% and fake rate of 1\%, see e.g.~\cite{ATLAS:2012ima}.

Our narrow jets tend to be quite hard since we are interested in the high-$m_{t \bar t}$ region but we have checked that the leading narrow jet $p_T$ distribution peaks in the range from 50 GeV to 300 GeV where the MV1 algorithm used by ATLAS outperforms this naive estimate~\cite{Aad:2015ydr} for our signal samples. To reflect the degradation of performance at higher $p_T$, we use a fake rate for light quarks and gluons of 10\% above 300 GeV. We have checked that combining the $p_T$-dependent $b$-tagging with contemporary top-tagging techniques renders the $Wjj$ background negligible compared to SM $t \bar{t}$ production at our signal mass points. We expect other SM backgrounds to be negligible: we find lower signal Acceptance $\times$ Efficiencies than the 13 TeV ATLAS study in \cite{ATLAS-CONF-2016-014} thanks to our stricter top-tagging which further suppresses all non-$t\bar t$ backgrounds. The final sensitivity of our study could potentially be improved by using a more permissive top-tagging algorithm and taking care to estimate non-$t \bar t$ background contributions.

In the next step, we require missing transverse energy $\slashed{p}_T$ with $|\slashed{p}_T| > 20$ GeV and $|\slashed{p}_T| + m_T > 60$ GeV where $m_T = \sqrt{2 p_{T,l} |\slashed{p}_T| (1-\cos \phi_{l \slashed{p}})}$.

\begin{figure}[t!]
\includegraphics[width=0.45\textwidth]{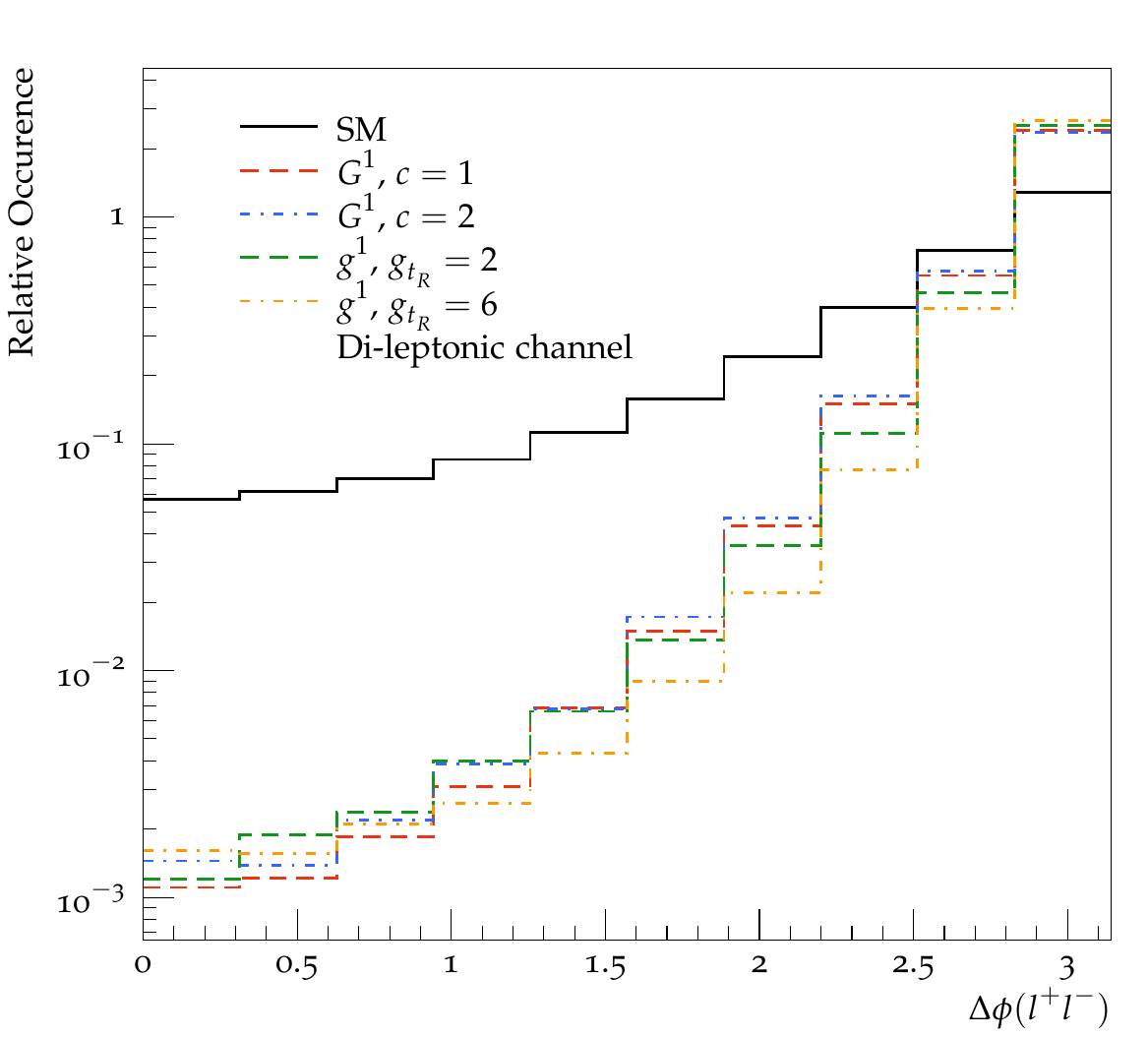}
\caption{Distribution of $\Delta \phi (l^+ l^-)$ for the considered scenarios for invariant masses $m(t \bar t) > 2$ TeV.}
\label{fig:dphill}
\end{figure}

We reconstruct the leptonic $W$ by assuming that its decay products are the leading lepton and a neutrino, which accounts for all of the reconstructed missing transverse momentum. The longitudinal component of the neutrino momentum is found by assuming the $W$ is produced on-shell, and we choose between the two resulting solutions by picking the one which minimizes $|m_{bl\nu} - m_t|$ after combining with the leading $b$-tagged jet. This object is our leptonic top candidate.

We extract $m(t \bar{t})$ by adding the found leptonic and hadronic top candidates and define $\theta_{l^{\pm}}$ by boosting to the leptonic top's rest frame and taking the angle between the lepton and the top's direction of travel.\footnote{Note that here are studies~\cite{Krohn:2009wm} that aim to extract the polarisation information from boosted hadronic tops but we do not attempt to do so here. We can expect the sensitivity of such a measurement to be smaller than that of the leptonic side measurement.}

\begin{figure*}[t!]
\centering
\subfigure[Di-leptonic channel.\label{fig:dileptonanglemass3dl}]{\includegraphics[width=0.45\textwidth]{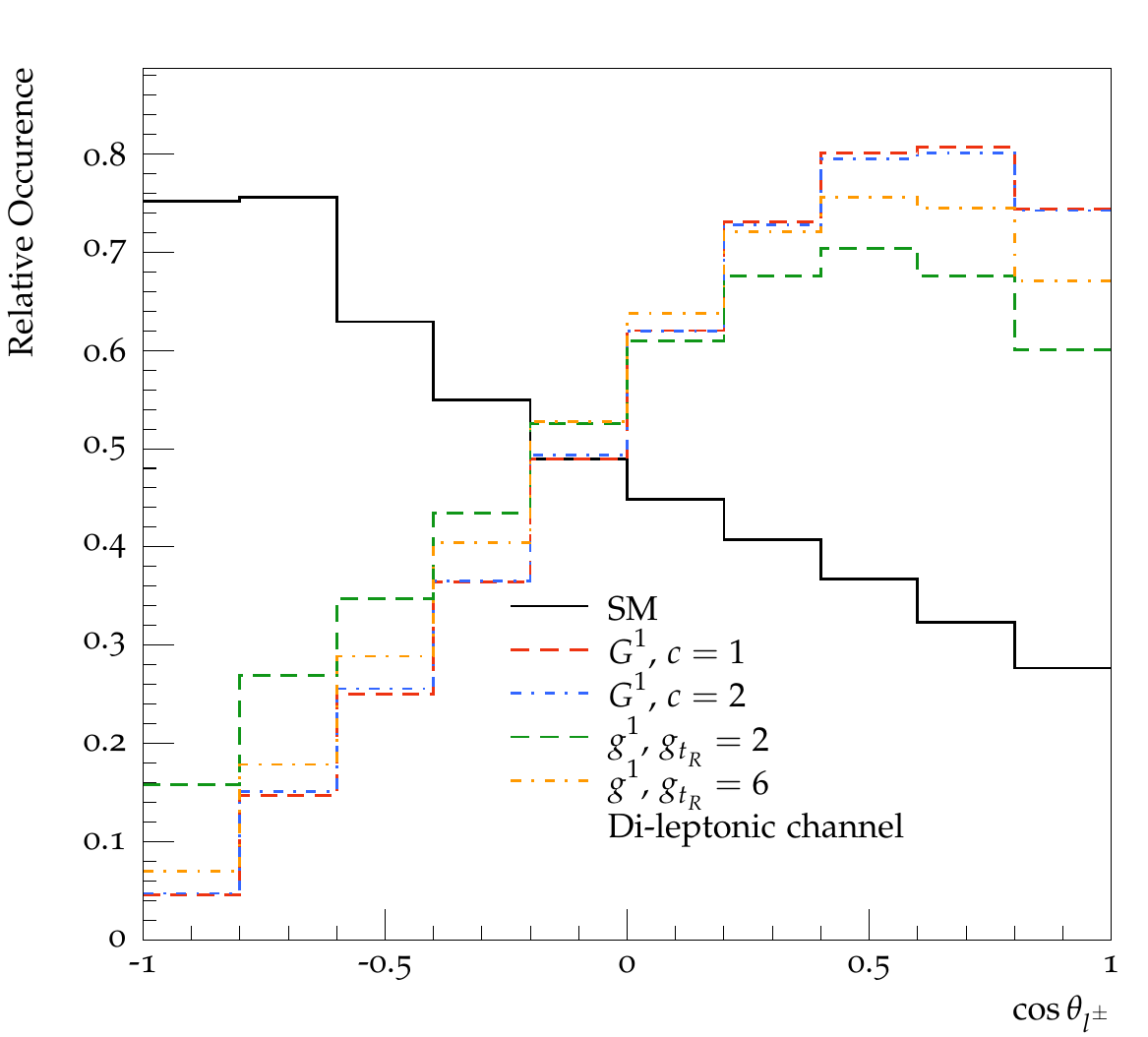}}
\hfill
\subfigure[Semi-leptonic channel.\label{fig:dileptonanglemass3sl}]{\includegraphics[width=0.45\textwidth]{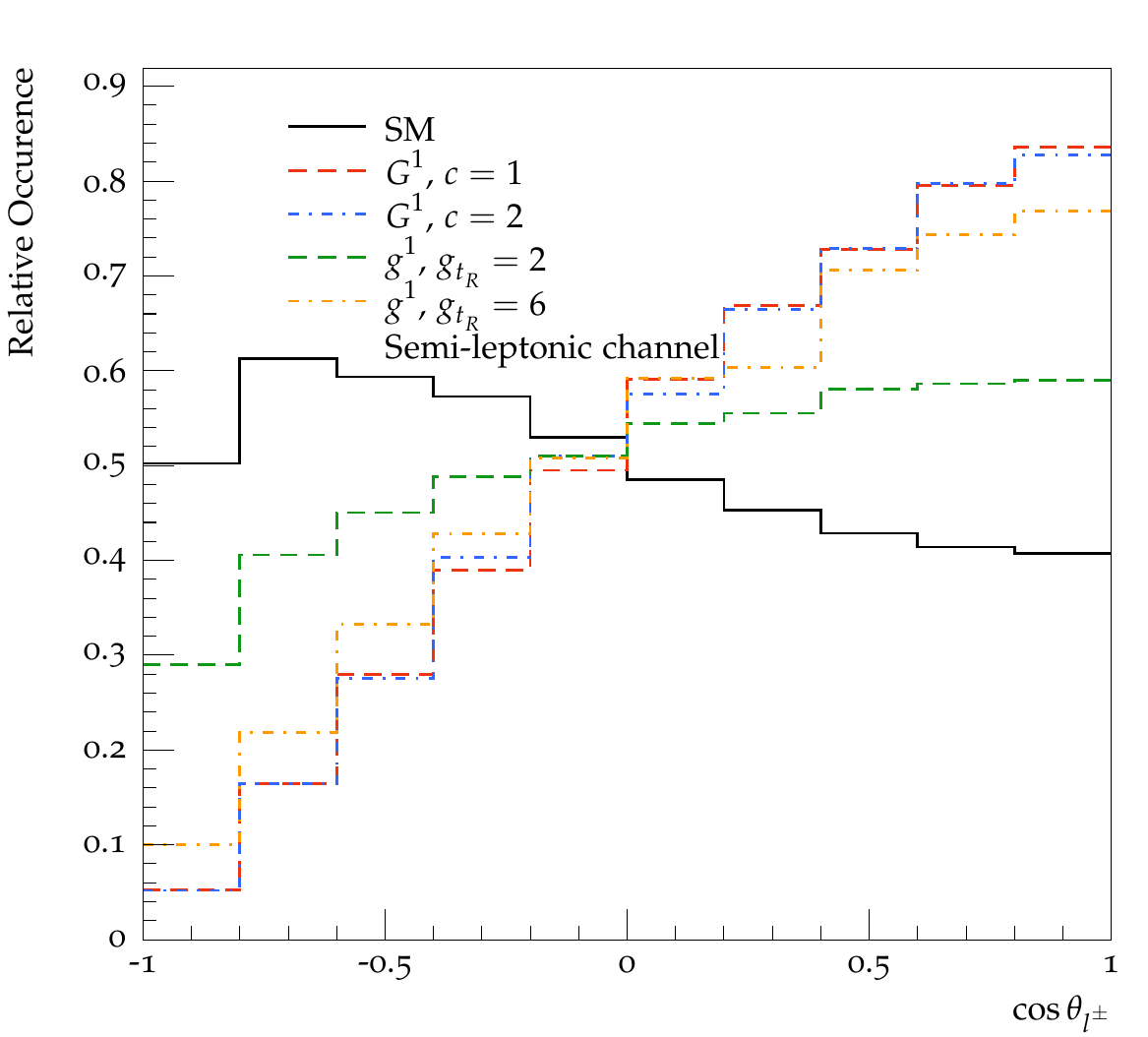}}
\caption{ $\cos \theta_{l^{\pm}}$ distributions for the SM $t\bar{t}$ and signal samples for the di-leptonic (a) and semi-leptonic (b) analyses, in both for $m(t \bar t) > 2$ TeV. Since the signal produces right-handed tops we see a large modification of these lepton angle distributions when compared to the SM expectation which at these high invariant masses is slightly left-handed. Note that the polarisation of the tops from $g^1$ decays differs between the two coupling choices and this can be discerned in both analyses.}
\label{fig:dileptonanglemass3}
\end{figure*}

\subsection{Di-leptonic study}
\label{sec:dilep}

The semi-leptonic final state discussed in Sec.~\ref{sec:semilep} is naively much more attractive due to a six times larger branching fraction (since we are only considering electrons and muons) and a less involved reconstruction of the individual top momenta. Nonetheless, it is worthwhile to also consider the di-leptonic final state as it offers two clean final state leptons which enable a comparably straightforward measurement of spin correlations with increasing statistics.

When considering di-leptonic $t \bar{t}$ decays, however, we run into a qualitatively new issue related to the reconstruction of the individual top momenta: with two neutrinos in the final state, we will have to make an educated guess of how the single missing transverse energy vector decomposes into the transverse components of the neutrinos $p_{T,\nu/\bar \nu}$ before reconstructing the longitudinal momentum components. There are a number of approaches that we outline in the following.

\begin{figure*}[t!]
\centering
\subfigure[\label{fig:semileptonanglemassdl}]{\includegraphics[width=0.45\textwidth]{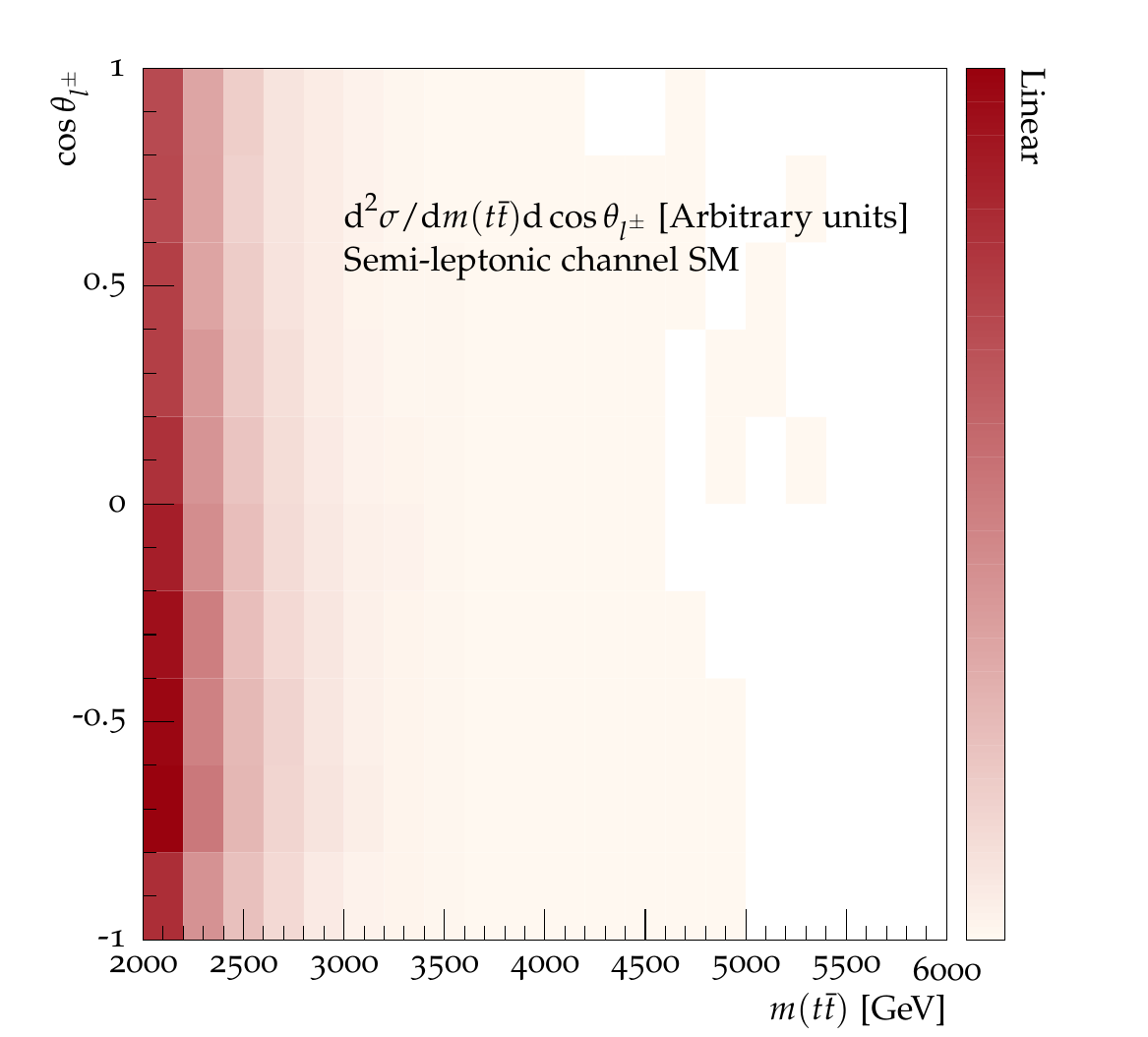}}
\hfill
\subfigure[\label{fig:semileptonanglemasssl}]{\includegraphics[width=0.45\textwidth]{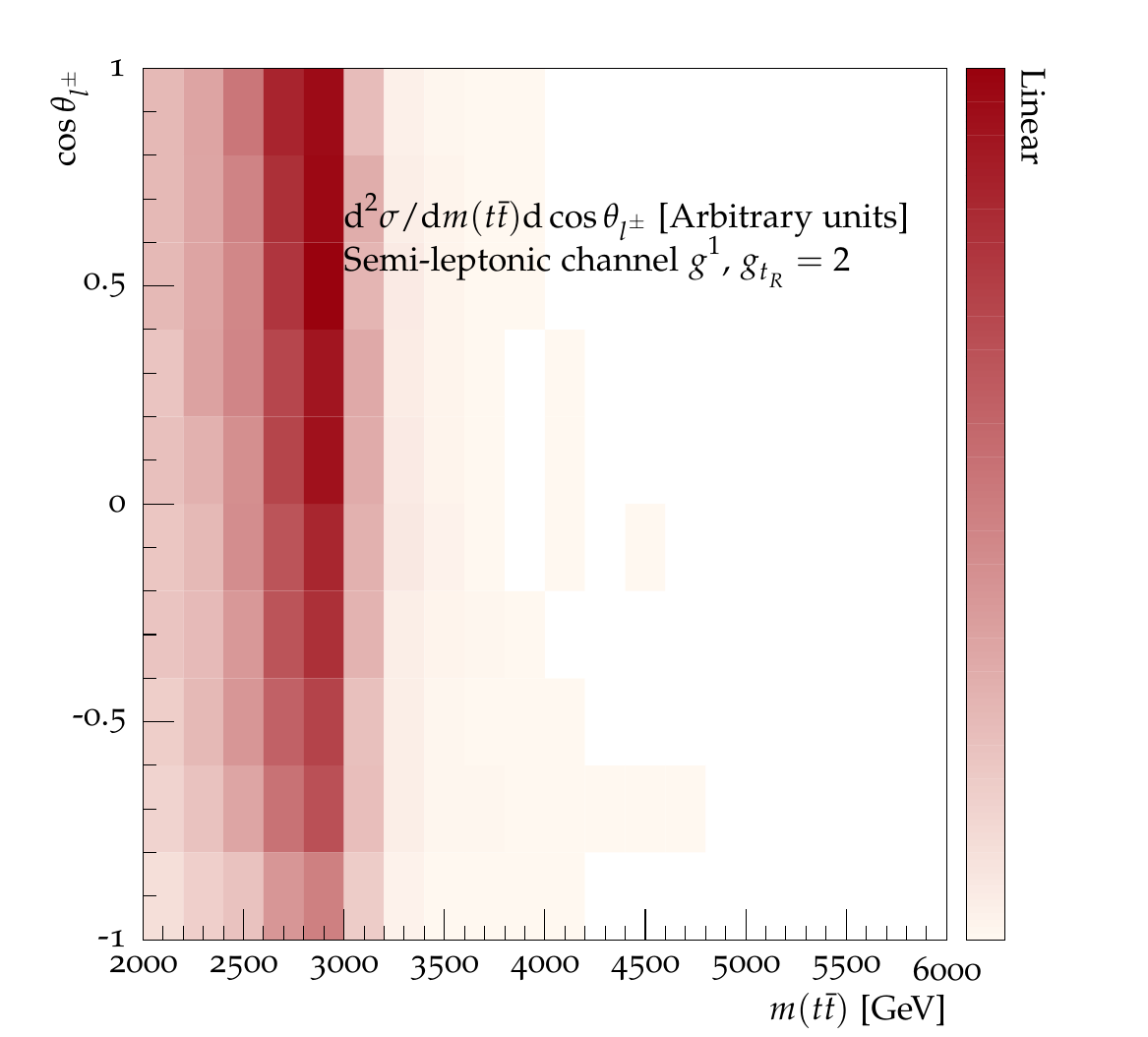}}
\caption{Two-dimensional shape distributions of $m(t \bar{t})$ and $\cos \theta_{l^{\pm}}$ for the expected SM background (a) and a narrow ($g_{t_R}=2$) $g^1$ (b) in the semi-leptonic analysis.}
\label{fig:semileptonanglemass}
\end{figure*}

\begin{figure*}[t!]
\centering
\subfigure[\label{fig:dileptonanglemass2dl}]{\includegraphics[width=0.45\textwidth]{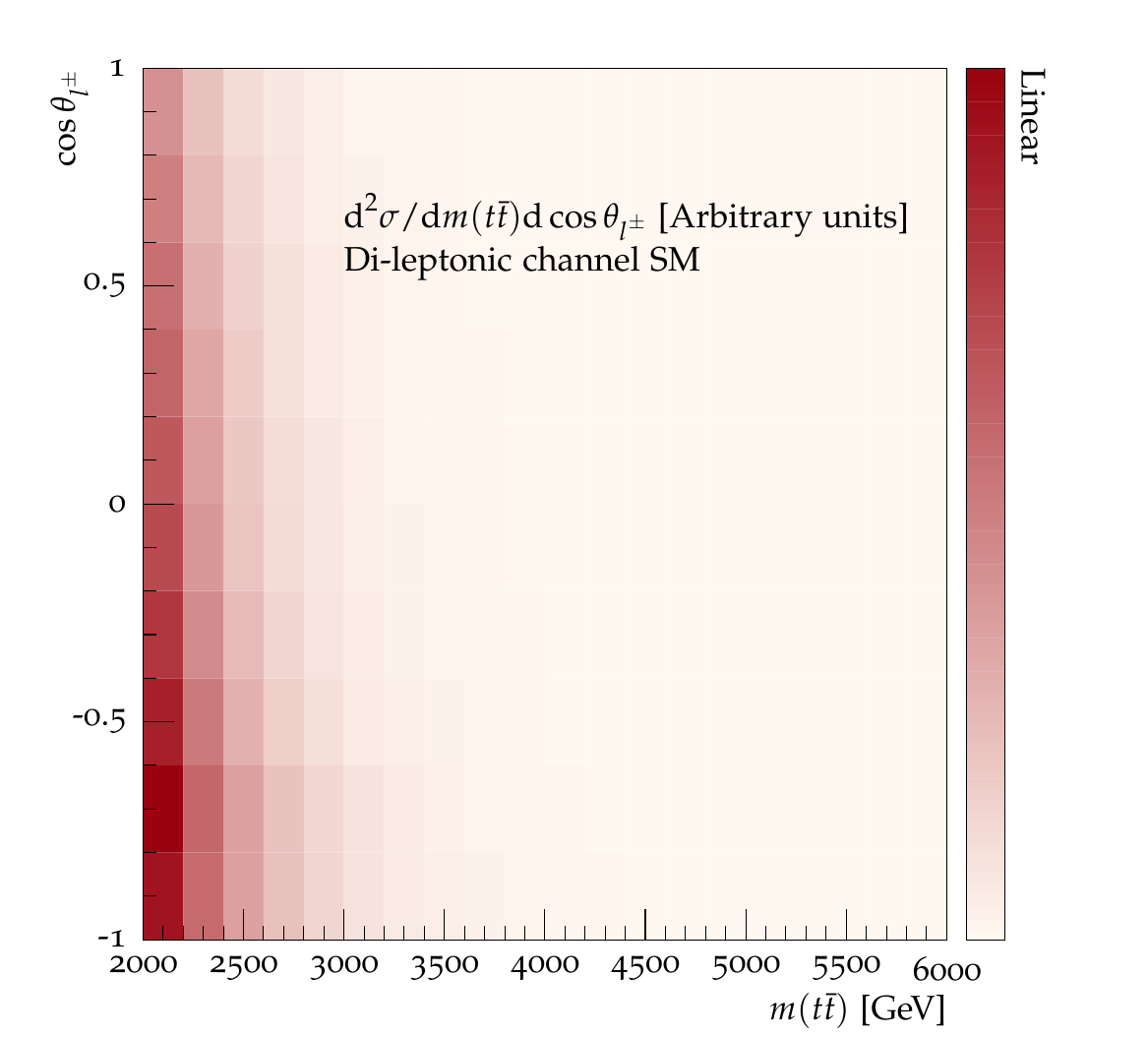}}
\hfill
\subfigure[\label{fig:dileptonanglemass2sl}]{\includegraphics[width=0.45\textwidth]{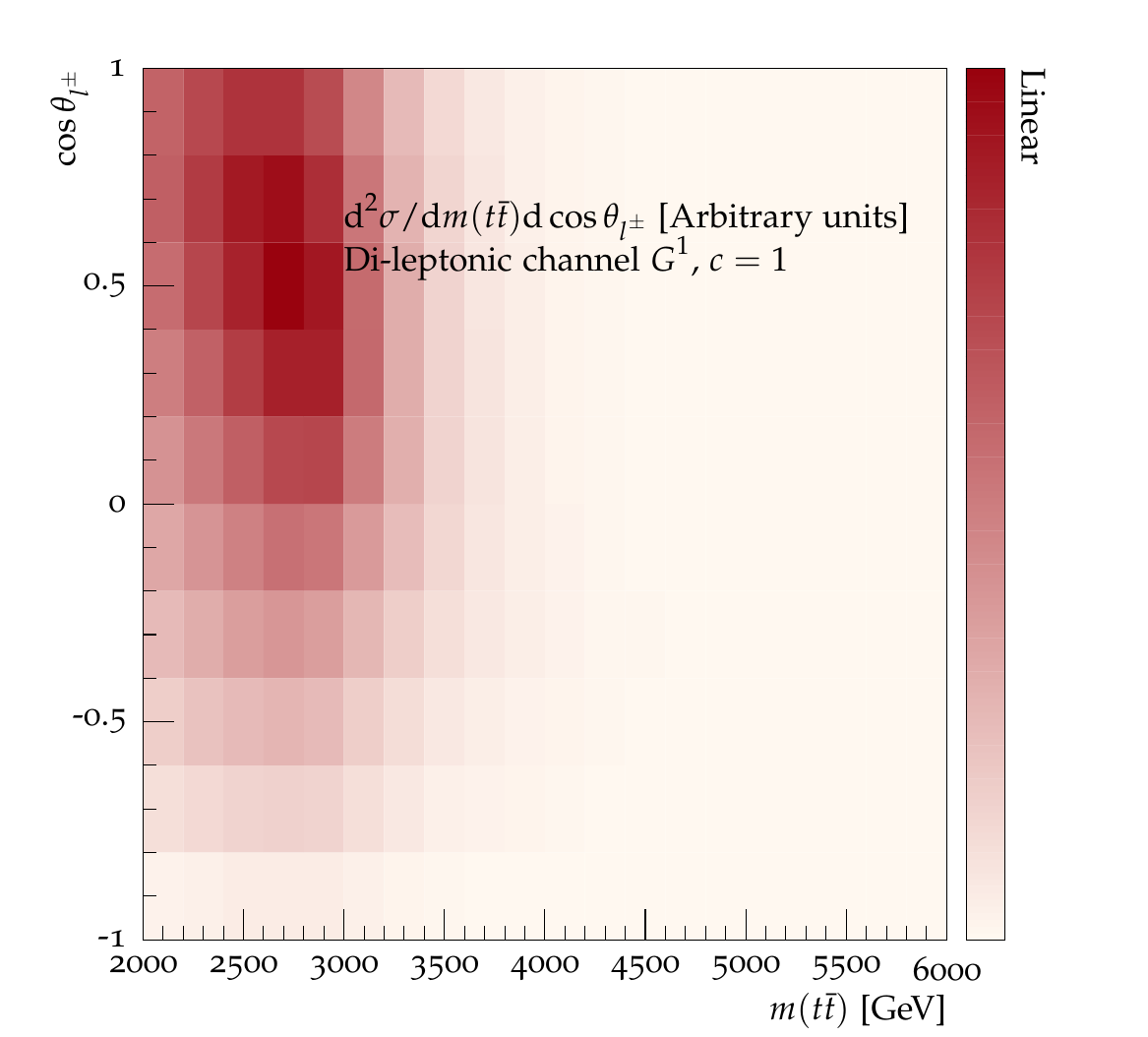}}
\caption{Two-dimensional shape distributions of $m(t \bar{t})$ and $\cos \theta_{l^{\pm}}$ for the expected SM background (a) and a narrow ($c=1$) $G^1$ (b) in the di-leptonic analysis. }
\label{fig:dileptonanglemass2}
\end{figure*}

The first method is to simply solve the full system of kinematic equations by assuming all intermediate particles are produced on-shell and that your measured kinematic quantities are exact~\cite{Sonnenschein:2005ed,Sonnenschein:2006ud}. This will in general provide up to eight sets of solutions, one of which being close to the true momenta assuming that the assumptions are valid. Using smeared kinematic quantities results in a larger mean number of solutions which causes large combinatorial uncertainties. CMS have made use of this approach together with a Matrix Element-method~\cite{Kondo:1988yd} to reduce the number of solutions on the basis of the matrix element weight.

A second method is to use so-called ``neutrino weighting'' \cite{Abbott:1997fv,Aaboud:2016syx}, which scans over a large number of proposed neutrino solutions and constructs and assigns individual weights for each guess based on how well the solution solves the kinematic equations. It is then possible to calculate observables for single events by either selecting the solution with the highest weight, or adding up the values for all solutions with correct weighting. This method is often used by ATLAS and has the advantage of only relying on kinematic information.

\begin{figure*}[t!]
\centering
\subfigure[\label{fig:c1sleplimit}]{\includegraphics[width=0.45\textwidth]{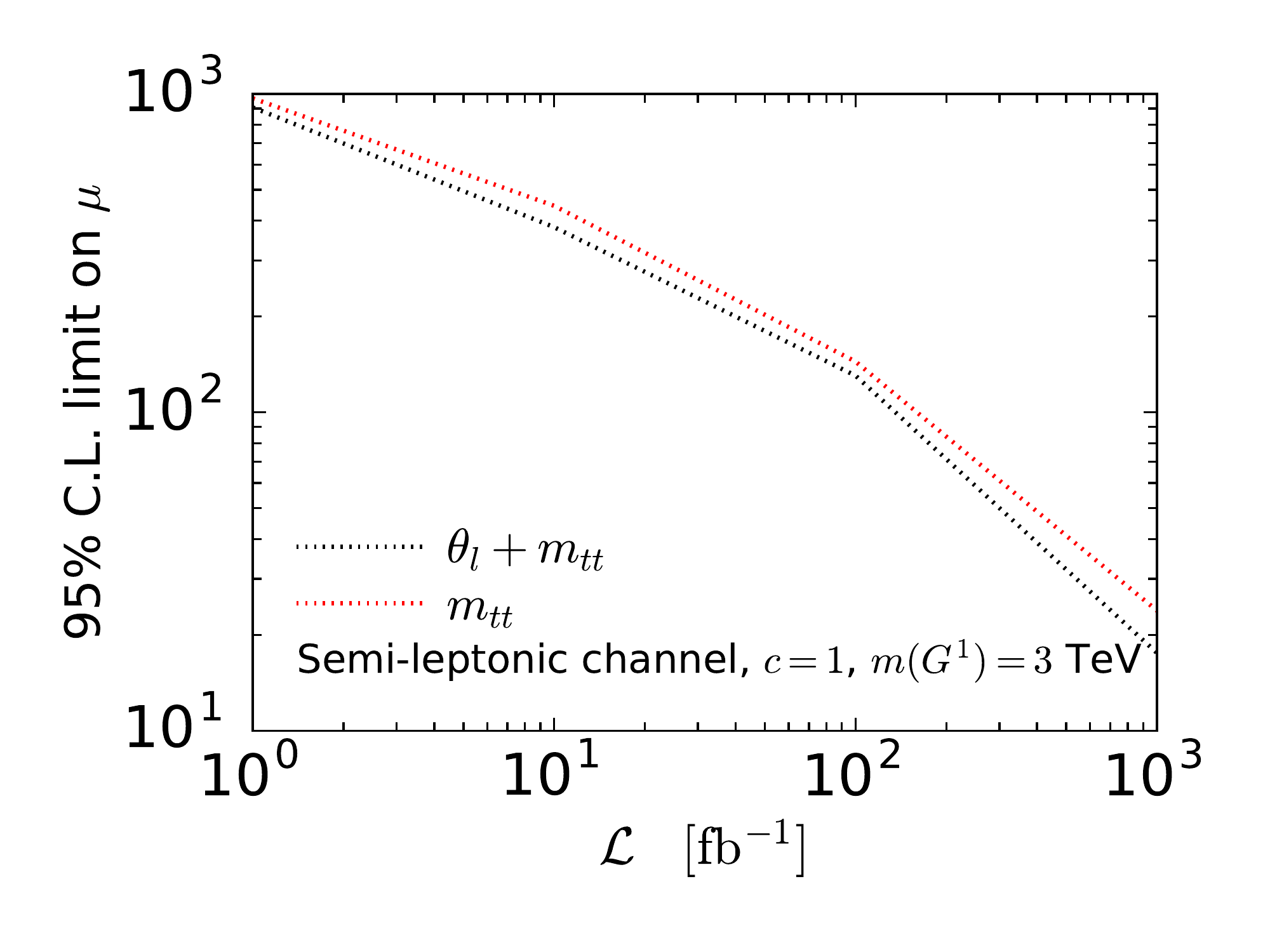}}
\hfill
\subfigure[\label{fig:g6sleplimit}]{\includegraphics[width=0.45\textwidth]{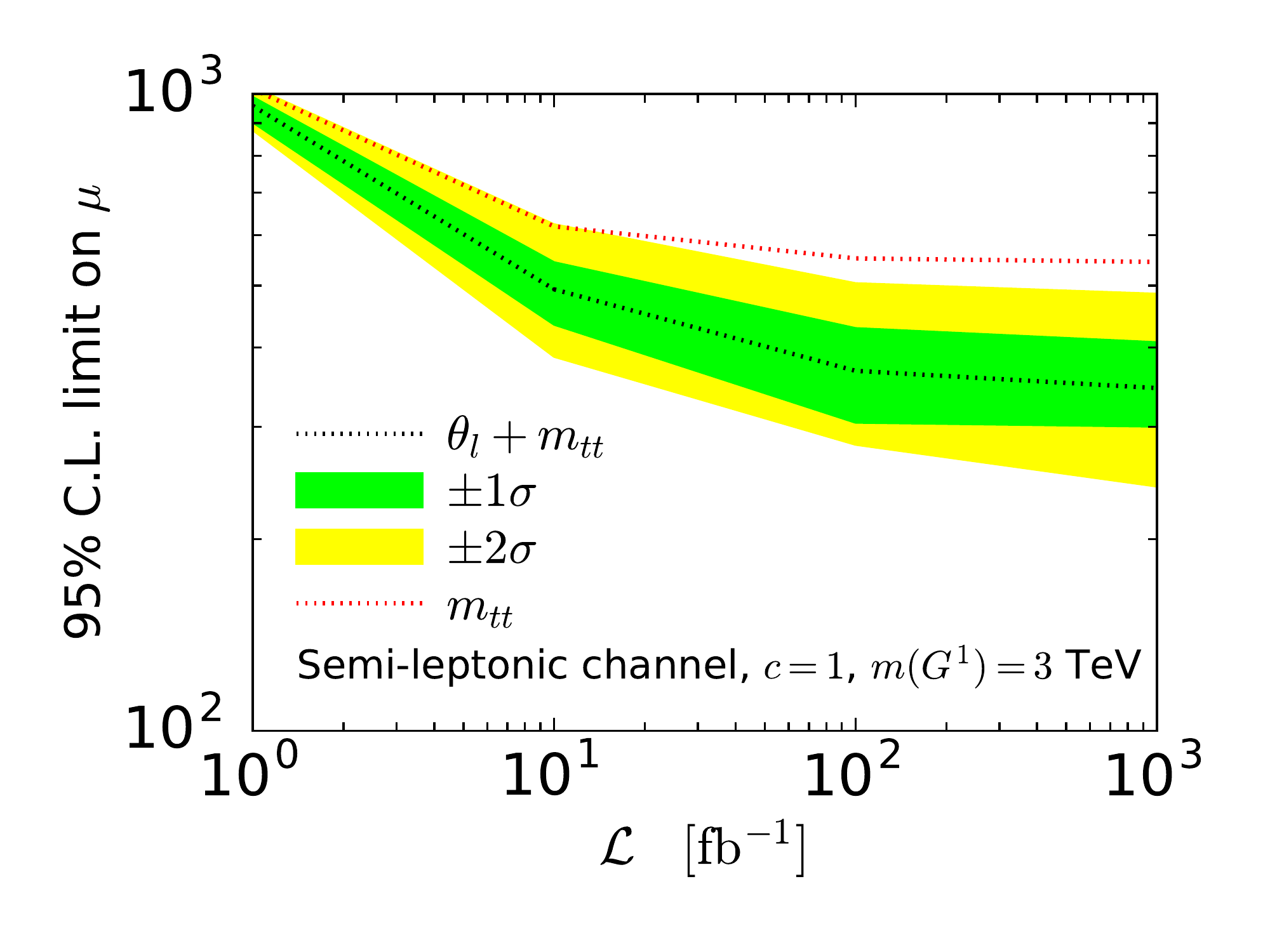}}
\caption{Limits on $\mu$ for a narrow ($c=1$) $G^1$ assuming (a) no systematics and (b) 5\% systematics (see text for details on how this is propagated to the individual bins) which can be set with different assumed total luminosities using $m(t \bar{t})$ and $\cos \theta_{l^{\pm}}$ (black line) and only using $m(t \bar{t})$ (red line) with the semi-leptonic analysis. The $\pm \sigma$ bands are for the combined result. }
\label{fig:sleptonlimit}
\end{figure*}

\begin{figure*}[t!]
\centering
\subfigure[\label{fig:c1dileplimit}]{\includegraphics[width=0.45\textwidth]{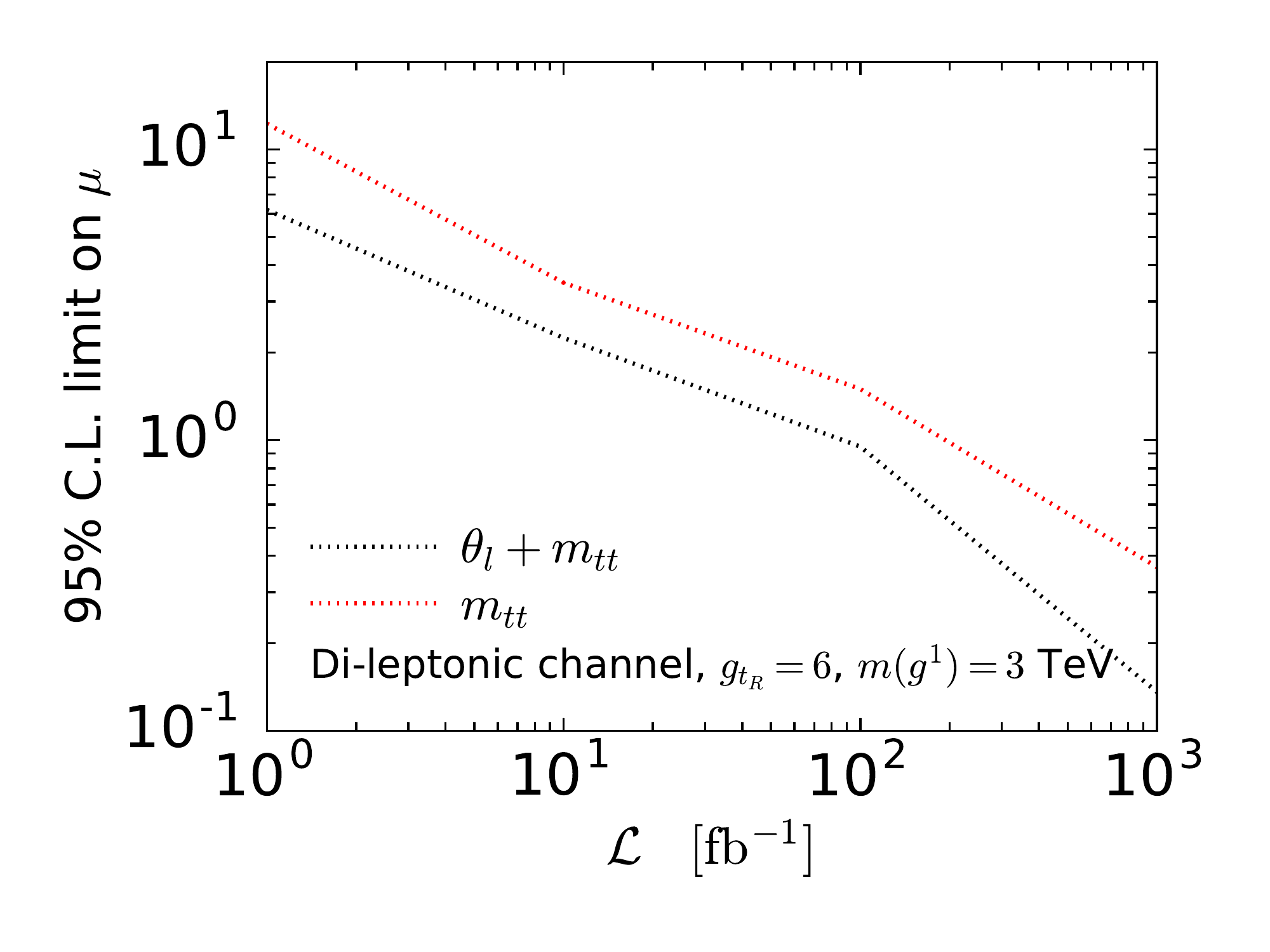}}
\hfill
\subfigure[\label{fig:g6dileplimit}]{\includegraphics[width=0.45\textwidth]{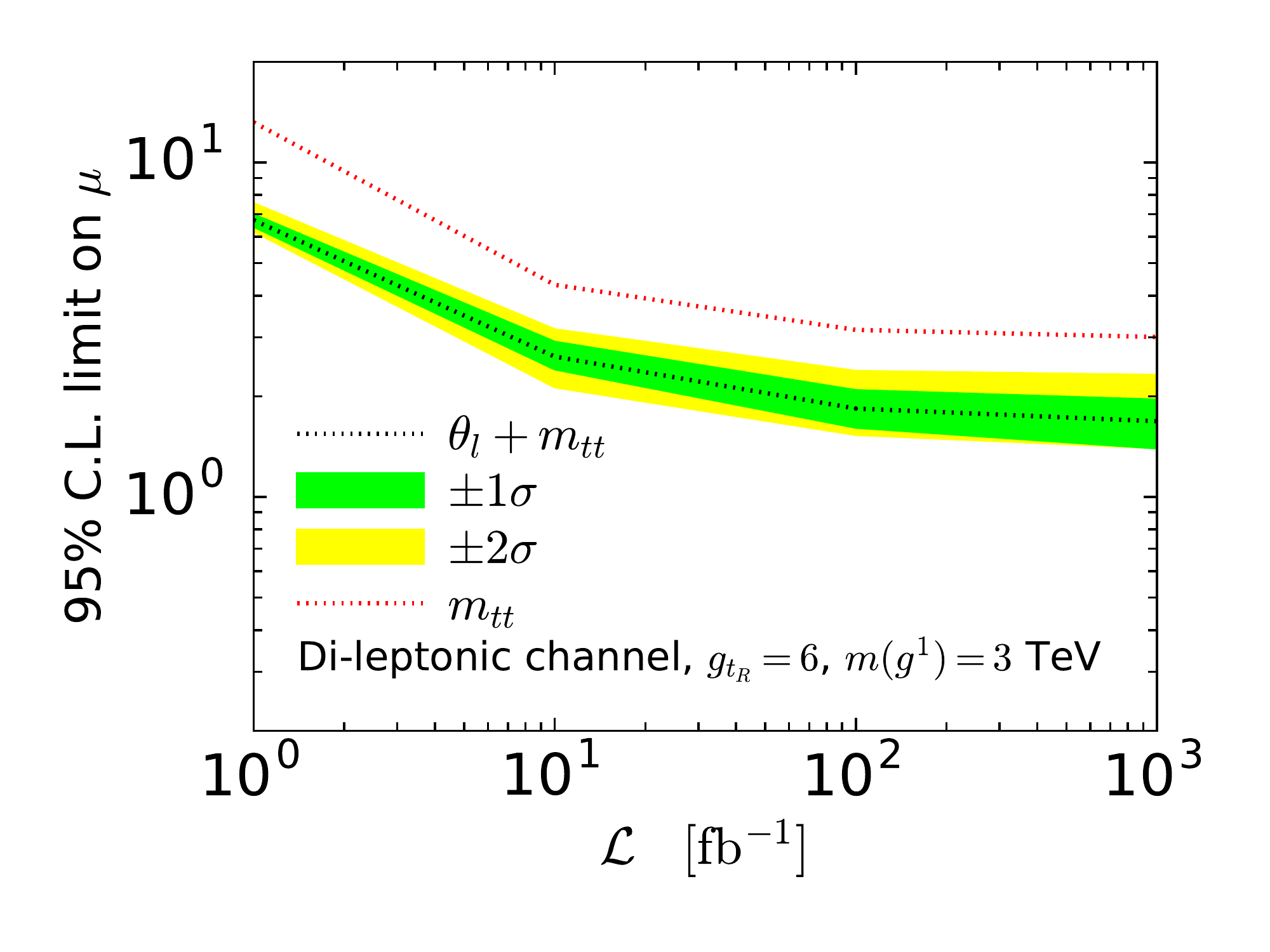}}
\caption{Limits on $\mu$ for a wide ($g_{t_R} = 6$) $g^1$ assuming (a) no systematics and (b) 5\% systematics on the total cross section (see text for details on how this is propagated to the individual bins) which can be set with different assumed total luminosities using $m(t \bar{t})$ and $\cos \theta_{l^{\pm}}$ (black line) and only using $m(t \bar{t})$ (red line) with the di-leptonic analysis. The $\pm \sigma$ bands are for the combined result.}
\label{fig:dileptonlimit}
\end{figure*}

A third method, which is the one we will adopt in this work, uses kinematic insights from the $M_{T2}$~\cite{Lester:1999tx} observable. The so-called $M_{T2}$ Assisted On Shell (MAOS) method~\cite{Cho:2008tj,Cho:2009wh} uses the solution for the transverse components of the two neutrino momenta which provides $M_{T2}$.  The bisection method for calculating $M_{T2}$ \cite{Cheng:2008hk} and subsequent improvements of the algorithm~\cite{Barr:2009jv,Lester:2014yga,Etayo2006324} have made it possible to find the solution numerically. The solutions for the neutrino momenta $k^{\pm}_{\nu/\bar\nu}$ (where $\pm$ denotes the remaining twofold ambiguity in the longitudinal components) will approach the true solutions for $M_{T2} \to m(t)$, with $k^{\pm}_{\nu/\bar\nu} = p_{\nu/\bar \nu}$\footnote{In this very particular situation we should find $k^+ = k^-$.} for $M_{T2} = m(t)$ with all kinematic quantities measured exactly and all intermediate particles on-shell. Therefore this approach provides an approach to improve the quality of the reconstruction if required by only using events with $m(t) - M_{T2} < C$ for some cut $C$.

\subsubsection{Analysis Selections and Reconstruction}

We begin the analysis by finding electrons with $p_T > 25$ GeV inside $|\eta| < 2.47$ and muons with $p_T > 25$ GeV inside $|\eta| < 2.7$. We then find anti-$k_T$ $R=0.4$ jets with $p_T > 25$ GeV with $|\eta| < 2.8$. Again we have to accept some overlap between the leptons and jets due to the large top boost, so we do not require these to be isolated and again assume we can separate very hard prompt leptons from a nearby jet. We then $b$-tag the jets within $|\eta| < 2.5$ with 70\% efficiency and a 1\% fake rate (10\% for $p_T > 300$ GeV with the comments regarding this choice made in Sec.~\ref{sec:semilep-analysis} also valid here), and require at least two $b$-tags. We also require missing transverse energy $\slashed{p}_T$ with $|\slashed{p}_T| > 60$ GeV.

\begin{figure*}[t!]
\centering
\subfigure[\label{fig:c1semillimitmass2}]{\includegraphics[width=0.45\textwidth]{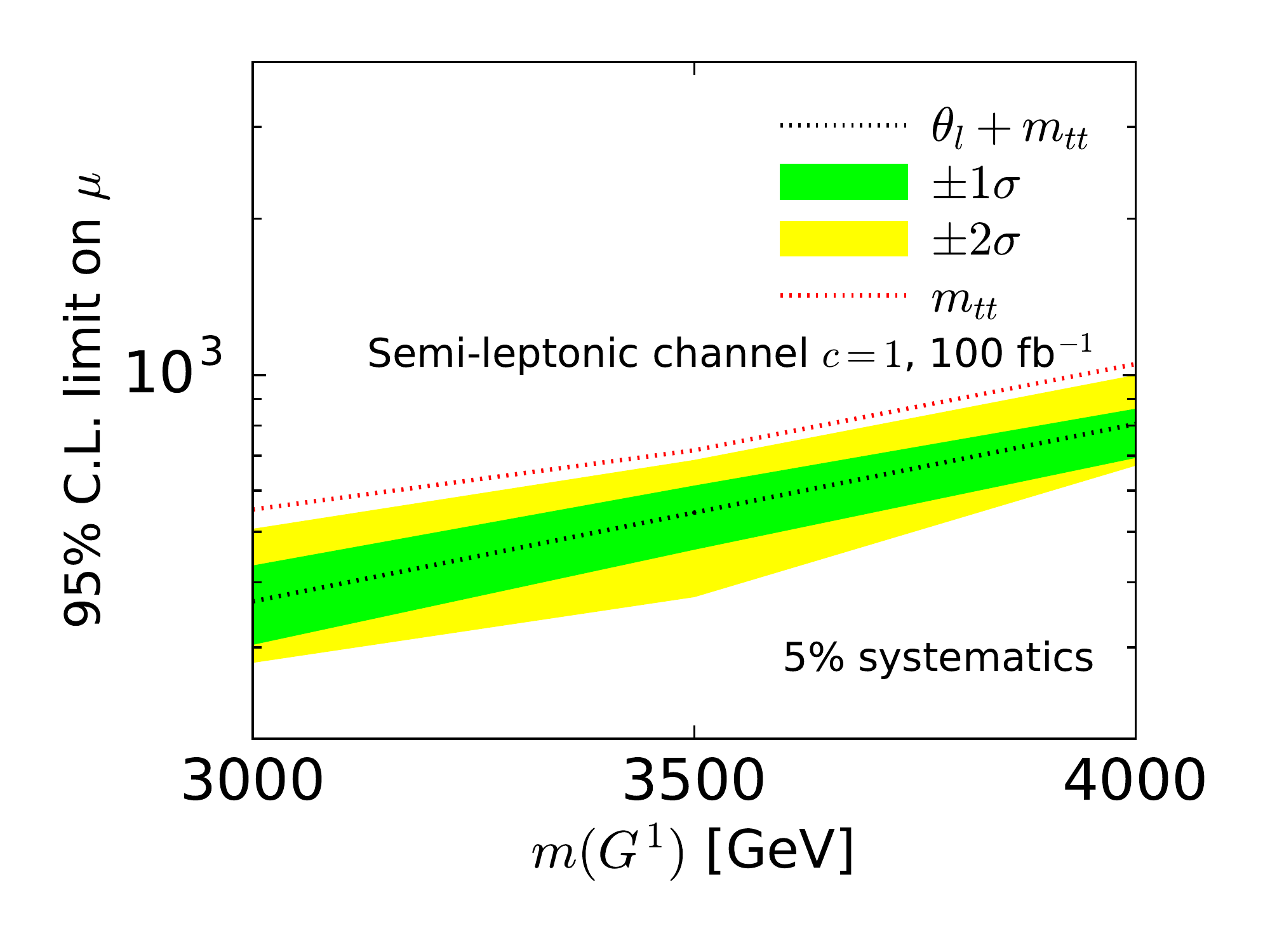}}
\hfill
\subfigure[\label{fig:c2semillimitmass2}]{\includegraphics[width=0.45\textwidth]{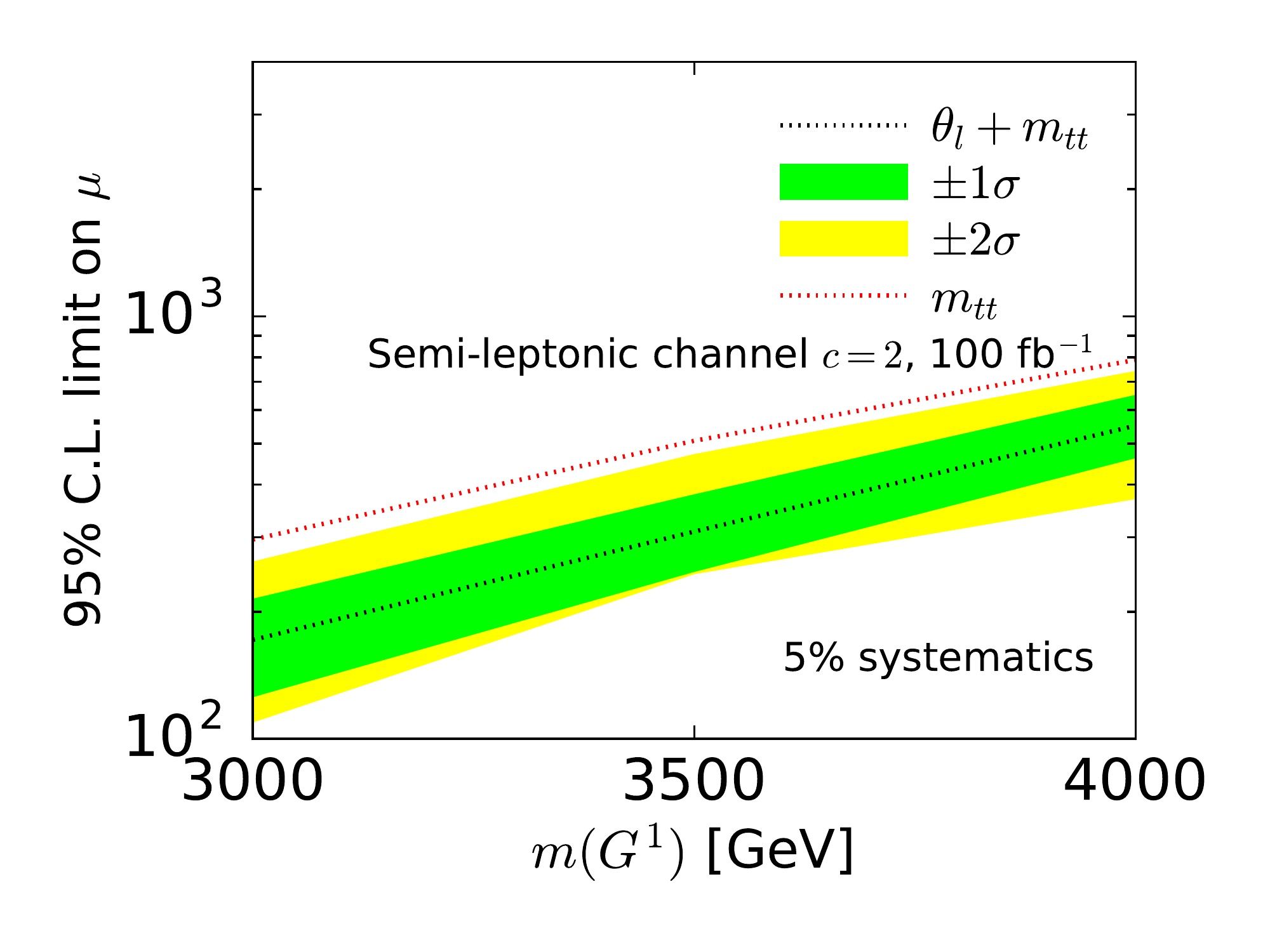}}
\caption{Limits on $\mu$ for (a) a narrow ($c=1$) $G^1$ and (b) a wide ($c=2$) $G^1$ for a fixed luminosity of 100 fb$^{-1}$ as a function of resonance mass using $m(t \bar{t})$ and $\cos \theta_{l^{\pm}}$ (black line) and only using $m(t \bar{t})$ (red line) with the semi-leptonic analysis. The $\pm \sigma$ bands are for the combined result.}
\label{fig:sleptonlimit2}
\end{figure*}

While the high boost of our tops means that we can usually correctly pair $b$-jets to leptons by taking the ones closest to each other in $\eta - \phi$ space, we make use of some standard approaches to further reduce the combinatorial uncertainty. Due to the large boost we consider, we do not gain much from cutting on $M_T^{t\bar{t}}(0)$, which is often considered in the literature~\cite{Barr:2009mx,Tovey:2010de,Choi:2011ys,Park:2011uz}, where $M_T^{t\bar{t}}(0)$ is defined as the transverse mass of the entire $t \bar{t}$ system when $m_{\nu \bar\nu} = 0$:
\begin{equation}
 \left(M_T^{t\bar{t}}(0)\right)^2 = m_{\text{vis}}^2 + 2 \left( \sqrt{ |p_T|^2 + m_{\text{vis}}^2 } |\slashed{p}_T| + p_T \cdot \slashed{p}_T \right)
\end{equation}
We therefore select the candidate which minimises at least two out of three test variables: $T_2$, $T_3$, and $T_4$ defined in~\cite{Choi:2011ys}. These correspond to how well the solution corresponding to each pairing reconstruct the $W$ and top masses and the expected $M_{T2}$ distribution. If either of the pairings returns complex solutions for the neutrino momenta we automatically select the other one. Once we have selected a pairing we veto the event if $M_{T2} > m(t)$ or $m_{bl} > \sqrt{m(t)^2 - m(W)^2}$.\footnote{Ignoring smearing, finite width effects, and $\mathcal{O}(m_{b})$ corrections to $m_{bl}$ these correspond to unphysical solutions.} Note that we change the pairing algorithm defined in~\cite{Choi:2011ys} slightly: this is because we find that vetoing the entire event if neither pairing results in a viable-seeming solution suppresses the $WWjj$ background with little signal efficiency loss. We do not use $m_{bl}$ for determining the correct pairing (referred to as the $T_1$ test variable in~\cite{Choi:2011ys}) since this would make the total number of test variables even and it correlates strongly with $T_2$.

\begin{figure*}[t!]
\centering
\subfigure[\label{fig:c1dillimitmass2}]{\includegraphics[width=0.45\textwidth]{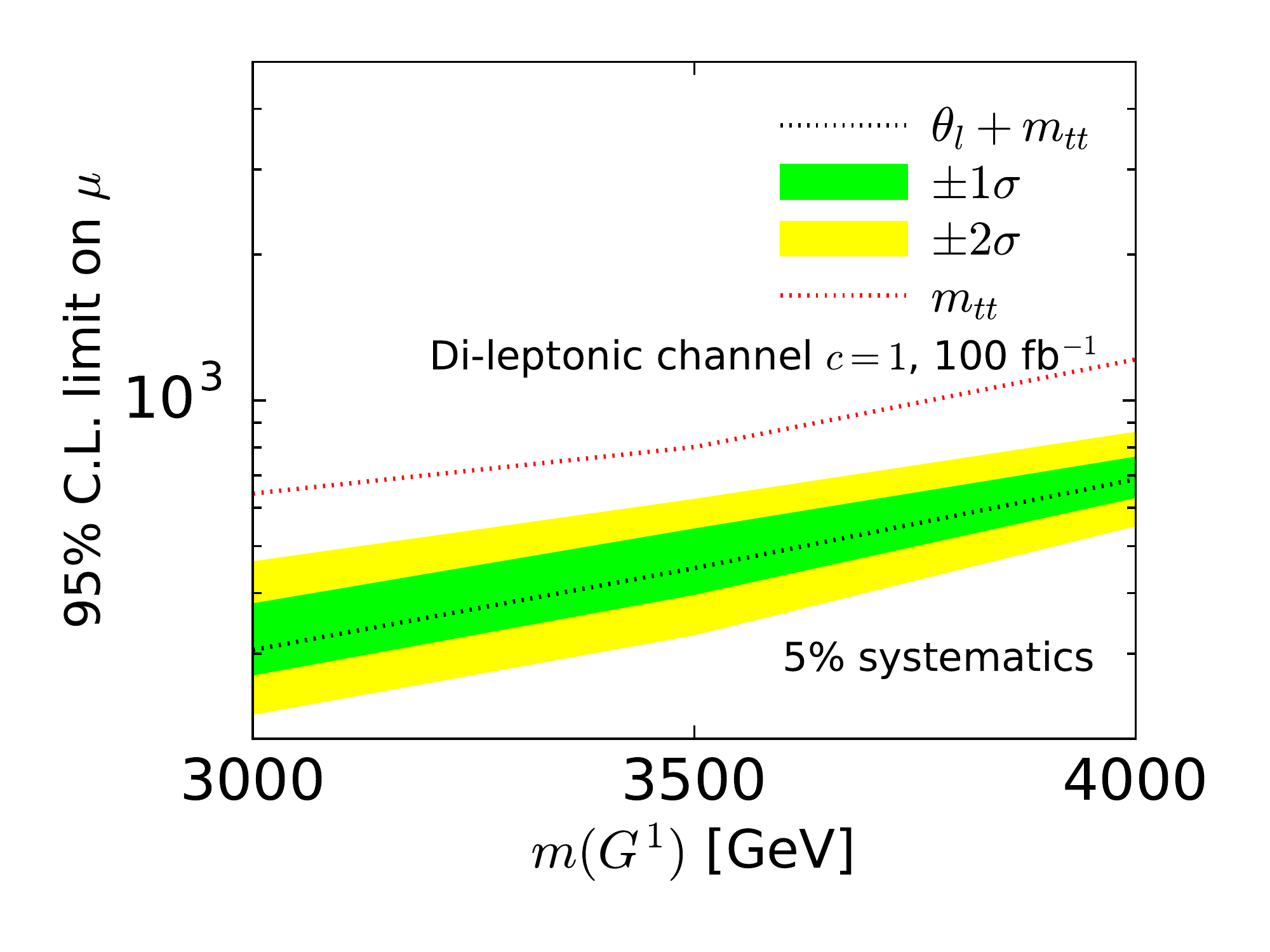}}
\hfill
\subfigure[\label{fig:c2dillimitmass2}]{\includegraphics[width=0.45\textwidth]{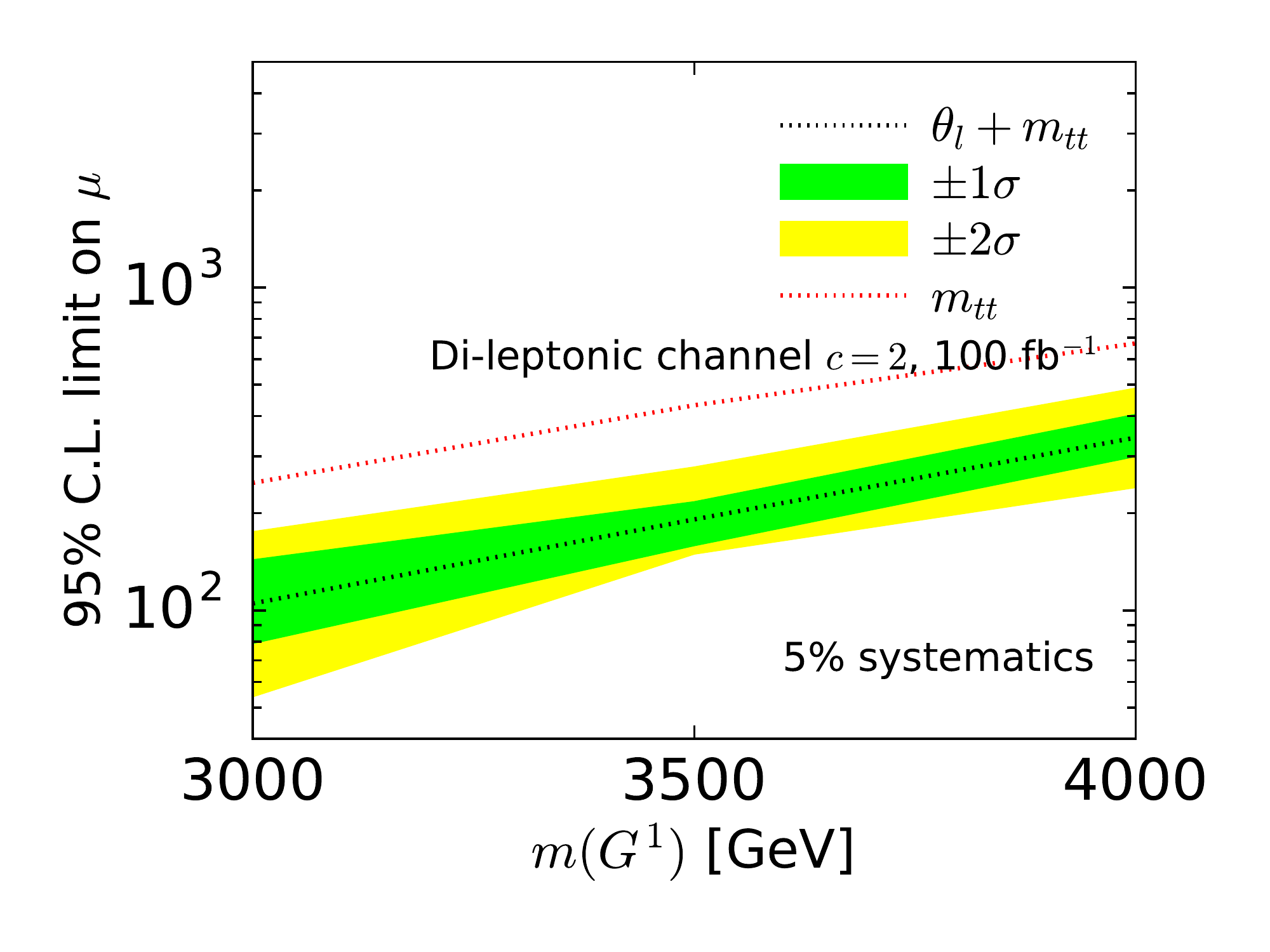}}
\caption{Limits on $\mu$ for (a) a narrow ($c=1$) $G^1$ and (b) a wide ($c=2$) $G^1$ for a fixed luminosity of 100 fb$^{-1}$ as a function of resonance mass using $m(t \bar{t})$ and $\cos \theta_{l^{\pm}}$ (black line) and only using $m(t \bar{t})$ (red line) with the di-leptonic analysis. The $\pm \sigma$ bands are for the combined result. }
\label{fig:dileptonlimit2}
\end{figure*}

%
\begin{figure*}[t!]
\centering
\subfigure[\label{fig:c1semillimitmass3}]{\includegraphics[width=0.45\textwidth]{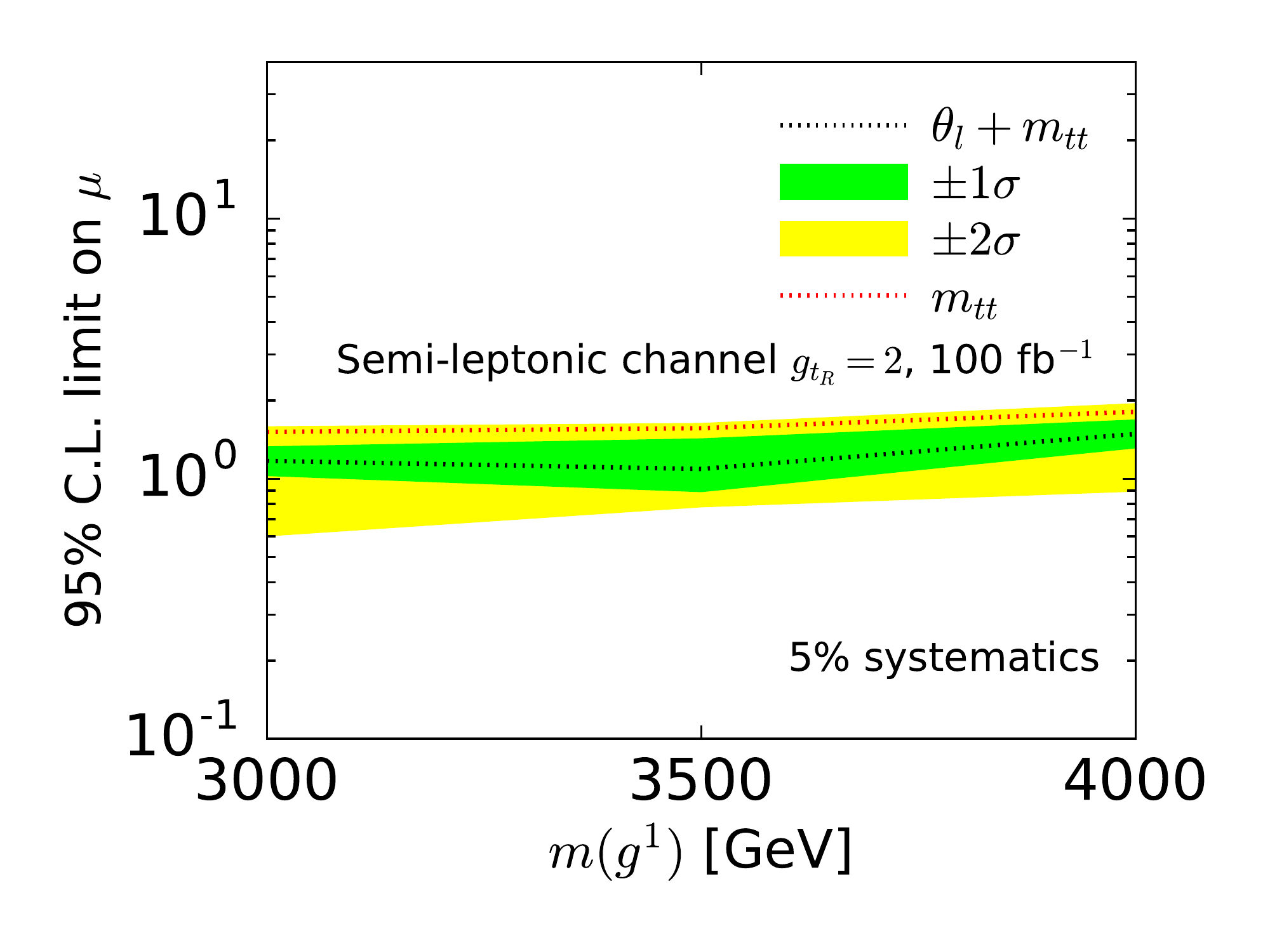}}
\hfill
\subfigure[\label{fig:c2semillimitmass3}]{\includegraphics[width=0.45\textwidth]{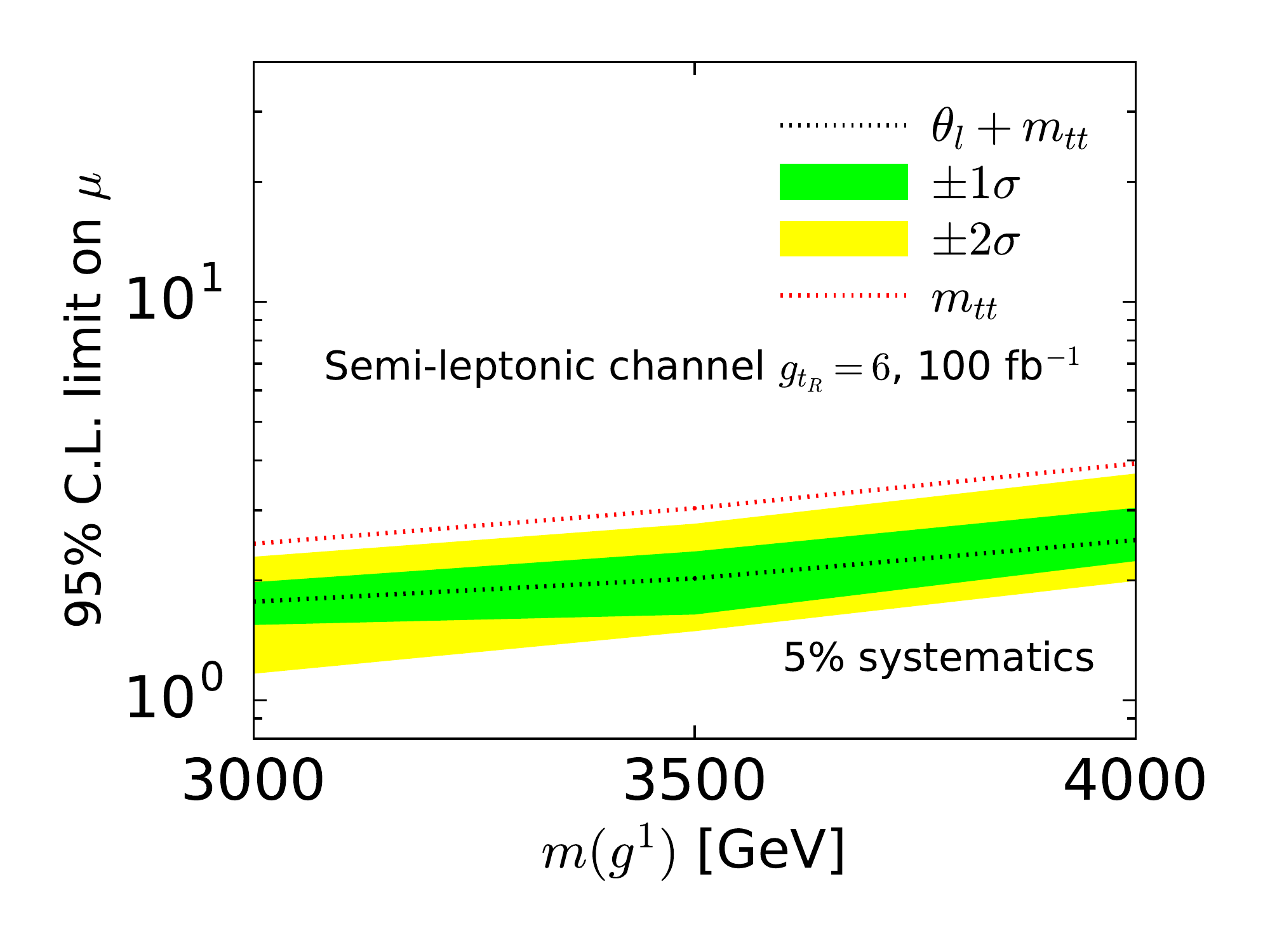}}
\caption{Limits on $\mu$ for (a) a narrow ($g_{t_R} = 2$) $g^1$ and (b) a wide ($g_{t_R} = 6$) $g^1$ for a fixed luminosity of 100 fb$^{-1}$ as a function of resonance mass using $m(t \bar{t})$ and $\cos \theta_{l^{\pm}}$ (black line) and only using $m(t \bar{t})$ (red line) with the semi-leptonic analysis. The $\pm \sigma$ bands are for the combined result.}
\label{fig:sleptonlimit3}
\end{figure*}

\begin{figure*}[t!]
\centering
\subfigure[\label{fig:c1dillimitmass3}]{\includegraphics[width=0.45\textwidth]{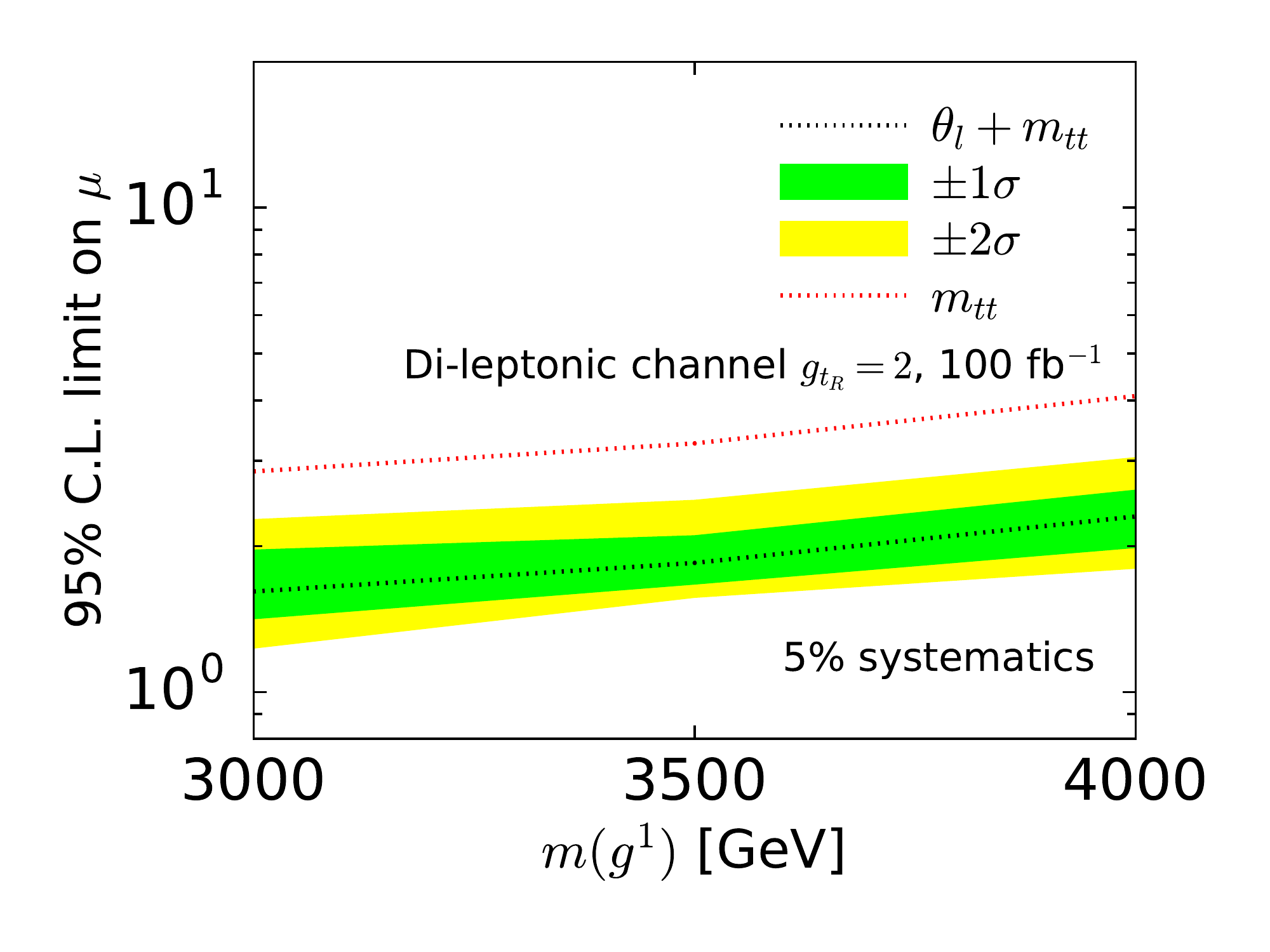}}
\hfill
\subfigure[\label{fig:c2dillimitmass3}]{\includegraphics[width=0.45\textwidth]{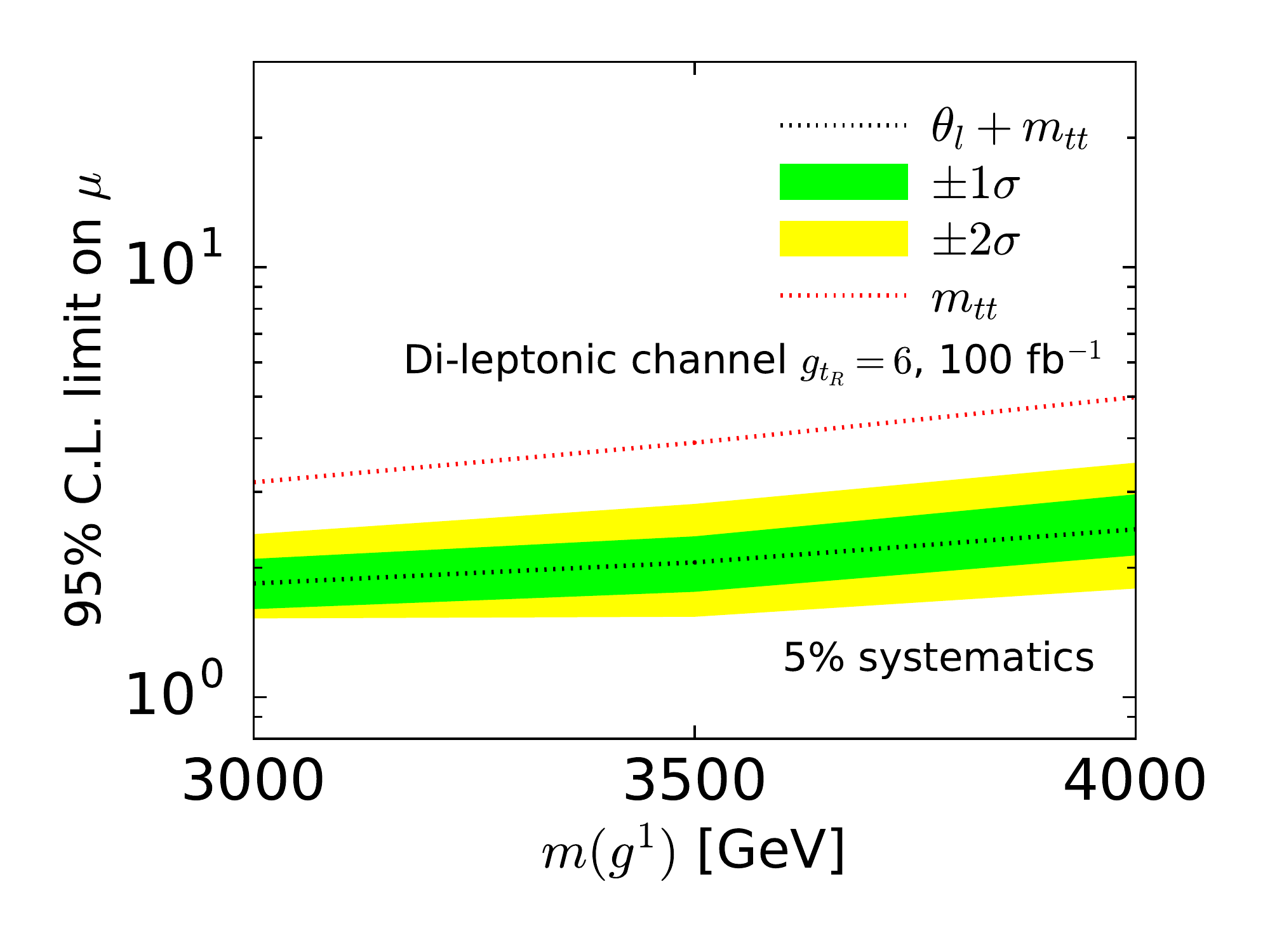}}
\caption{Limits on $\mu$ for (a) a narrow ($g_{t_R} = 2$) $g^1$ and (b) a wide ($g_{t_R} = 6$) $g^1$ for a fixed luminosity of 100 fb$^{-1}$ as a function of resonance mass using $m(t \bar{t})$ and $\cos \theta_{l^{\pm}}$ (black line) and only using $m(t \bar{t})$ (red line) with the di-leptonic analysis. The $\pm \sigma$ bands are for the combined result.}
\label{fig:dileptonlimit3}
\end{figure*}

As discussed above we reconstruct the individual neutrinos using the MAOS method.
We take the solution for the transverse momenta of the neutrinos which gives the correct $M_{T2}$
, and solve the remaining kinematic constraints to give two solutions for the longitudinal component of each neutrino. This results in four final solutions for the complete kinematics of the event with equal weights. This technique has been used for example in phenomenological studies of production angle measurements in~\cite{Cho:2008tj} and top polarisation measurements in~\cite{Guadagnoli:2013xia}.
Despite the fourfold combinatorial uncertainty, it reproduces truth-level angular observables well as this only affects the longitudinal neutrino momenta. However the mass resolution suffers greatly and as a result even narrow resonances end up as wide excesses rather than distinctive bumps in the mass spectrum. Unlike in the semi-leptonic case in Section \ref{sec:semilep-analysis} we can extract the lepton angle from both tops by again boosting to the individual rest frames and taking the angle of their decay lepton to the top direction of travel.

\section{Discussion of Results}
\label{sec:results}
\subsection{Signal vs Background discrimination}
We estimate the limits that can be set on the signal strength $\mu = \sigma/\sigma_\textrm{expected}$ for our model setups with the $m(t \bar{t})$ and combined $m(t \bar{t})$ - $\theta_{l^{\pm}}$ distributions by using the Modified Frequentist confidence level $CL_s$ as outlined in \cite{Junk:1999kv}: for each distribution we calculate the likelihood ratio
\begin{equation}
  X = \prod^{\textrm{bins}}_i e^{- \mu s_i} \left(1 + \frac{\mu s_i}{b_i} \right)^{d_i}
\end{equation}
where $s_i$, $b_i$ and $d_i$ are the expected number of signal and background, and observed number of events for each bin respectively. Using the likelihood ratio we can compute%
\begin{align}
  CL_{s+b} =& \hspace{0.05cm} P_{s+b} \left(X < X_\textrm{obs} \right)\,, \\
  CL_b   \hspace{0.35cm}  =& \hspace{0.05cm} P_{b}   \left(X < X_\textrm{obs} \right)\,, \\
  CL_s   \hspace{0.35cm}  =& \hspace{0.05cm} CL_{s+b} / CL_b\,.
\end{align}
To avoid spurious exclusions we do not use bins which have no background events -- this has a negligible effect as we have ensured there is sufficient statistics in all bins which are expected to contribute to the exclusion limit for our signal models.

A value of $CL_s < 0.05$ is interpreted as excluding the corresponding value of $\mu$ at 95\% confidence level~\cite{Read:2002hq}. While our statistical setup is meant to closely resemble those currently employed by the LHC experiments we would also advise interested readers to read the recent study in \cite{Ferreira:2017ymn} which investigates the information gain from using multi-dimensional distributions such as our $m(t \bar{t})$ - $\theta_{l^{\pm}}$ one using Bayesian methods.

When calculating limits we use a flat Gaussian systematic of 5\% on the total cross section\footnote{We can expect that data-driven methods, that use the low $m(t\bar t)$ spectrum to extrapolate to our signal region become well-controlled with large data sets.} of the background and only statistical uncertainties for the signal. To propagate the systematic uncertainty to individual bins we assume the fractional systematic error is the same in all bins, and calculate the correct uncertainty which would lead to the stated uncertainty on the total cross section when adding up all the bins assuming they are statistically independent. In general introducing systematic uncertainties and propagating these in a consistent manner always requires us to make an assumption of how this is to be done which introduces a large effect on the final limit on $\mu$. In order to provide an estimate of the importance of the systematic uncertainty on our limits we also present a comparison to limits calculated with no systematic uncertainties in Figs.~\ref{fig:sleptonlimit} and~\ref{fig:dileptonlimit}.

\subsection{Improvement from top polarisation observables}
Before we comment on the relative improvement from including polarisation-sensitive observables let us quickly investigate the expected phenomenology in the model we consider. As can be seen from Fig.~\ref{fig:ttbarmass}. The reconstruction smears out the resonance so the signal appears very wide for all signal models in the semi-leptonic and di-leptonic analysis. For relatively narrow resonances our reconstruction of the semi-leptonic channel yields a better performance, however, once moving to larger widths, the $m_{t\bar t}$ distribution quickly loses its peak-like features. In such a case, setting limits by using $m_{t\bar t}$ as a single discriminant effectively means constraining a continuum excess.

Considering directly-inferred angular quantities like $\Delta\phi(l^+l^-)$ from, e.g., the di-lepton final state does not offer a great discriminative power. This is in particular true when we would like to discriminate between different signal hypothesis once an excess has been discovered. The reason for the highly correlated $\Delta\phi(l^+l^-)$ is the large considered mass range of the $t\bar t $ resonance, which leads to back-to-back tops and leptons as a consequence.

It is exactly the boost to the top rest frame which lifts this degeneracy (modulo reconstruction inefficiencies). And since the signal produces highly polarised tops, we see a large modification of these lepton angle distributions, which provides additional discrimination power (Fig.~\ref{fig:dileptonanglemass3}) that we can use to tighten the estimated constraint on $\mu$ when combined with $t\bar t$, Figs.~\ref{fig:semileptonanglemass} and~\ref{fig:dileptonanglemass2} (we also show the distribution of the expected SM background which exhibits no particular resonant features in the $m(t\bar t)-\cos\theta_{l^\pm}$ plane).
Note that the polarisation of the tops from $g^1$ decays differs between the two coupling choices and this is visible in both channels.

Using the $m(t\bar t)-\cos\theta_{l^\pm}$ correlation as the baseline of the limit setting outlined above we obtain a large improvement by a factor up to $\sim 3$ with increasing luminosity compared to $m(t\bar t)$ alone in Fig.~\ref{fig:c2dillimitmass3} for the ideal case of the di-leptonic analysis of a wide highly polarised resonance, as the large statistics available with 100 fb$^{-1}$ provide an efficient sampling of the sensitivity unveiled in Figs.~\ref{fig:dileptonanglemass2}. This relative improvement reduces for smaller reconstructed widths that can be reached in the semi-leptonic channel as discriminating power in $m(t\bar t)$ is gained, yet an improvement at large luminosity by a factor of $\sim \sqrt{2}$ is still possible for our benchmark narrow less-polarised gluon in Fig.~\ref{fig:c1semillimitmass3}, which is the least sensitive of our parameter points.

It is exactly this improvement from including polarisation information, which renders the analyses potentially sensitive -- depending on systematics -- to broad gluon-like resonances at ${\cal{L}}\sim 100~\text{fb}^{-1}$ at our benchmark setting. Discrimination solely based on $m(t\bar t)$ flattens out and an analysis which focuses exclusively on resonant-like enhancements will have less sensitivity by factors up to 3.

The improvement is not too sensitive on the precise mass scale around our chosen benchmark, and becomes especially relevant at large widths as alluded to in the beginning of this work, Figs.~\ref{fig:sleptonlimit2}, \ref{fig:dileptonlimit2}, \ref{fig:sleptonlimit3}, and~\ref{fig:dileptonlimit3}.

As can be seen from our results for graviton-like resonances, depending on the size of the cross section, only including spin polarisation is not enough to reach a sensitivity to constrain the underlying model satisfactorily. Nonetheless the relative improvement by a factor of $\simeq 3$ should provide an important handle to tackle such low-cross section scenarios much better at large luminosity, possibly as part of a multivariate approach invoked by the experiments.

\section{Conclusions}
\label{sec:conclusions}
Resonance searches at the LHC $t\bar t$ final states are a well-motivated strategy for discovering new physics beyond the SM~\cite{Aad:2015fna,Khachatryan:2015sma}. While peaks in the mass spectrum are very powerful indicators of the presence of such new physics, we also often expect to see large modifications to other distributions and combining this information through multi-dimensional distributions often offers a good way to improve sensitivity. Additionally, if the resonance becomes wide, invariant mass distributions necessarily lose sensitivity. We have performed a detailed investigation of the semi-leptonic and di-leptonic $t\bar t$ final states for $\sqrt{s}=14~\text{TeV}$ and provide quantitative estimates of the information gain from including top polarisation information in the limit setting. Our results demonstrate that this information helps to ameliorate the loss in sensitivity for wider signal models. To make our analysis comparable to the practice of the experiments we have focussed on the RS scenario as a particular candidate that provides a theoretically well-defined framework for such a phenomenological situation. For the fully-polarised scenarios we study in this work we find improvements of factors of up to 3 (2) on the limit of the signal strength for the di-(semi)-leptonic analysis at large luminosity, with larger improvements for wider signal models as expected. For our benchmark choice of 3 TeV resonances, including this information is crucial to exclude gluon-like at 95\%. Interestingly the larger improvement for the di-leptonic analysis allows this channel to become competitive with semi-leptonic one for resonance searches for these types of models, however we would like to note that this statement heavily depends on the systematics modelling and only a dedicated experimental analysis can fully assess the relative sensitivities.

While these improvements are specific to our parameter choices at face value, similar relative improvements can be expected for other, non-graviton or gluon resonances (not limited to RS models) that predict a net polarisation of the top pair. Polarisation information is therefore an important ingredient to a more comprehensive analysis strategy that builds upon the invariant top pair mass, providing additional information in multivariate approaches.

\acknowledgement
We thank Simon Head for collaboration during an early stage of this work.
JF acknowledges support from the Helmholtz Gemeinschaft.
KN thanks the University of Glasgow College
of Science \& Engineering for a PhD scholarship.

\bibliography{references.bib}

\end{document}